\documentclass[11pt,a4paper]{article}
\pdfoutput=1
\usepackage{jheppub}
\usepackage{amsmath,amssymb,graphicx}
\usepackage{epsf}
\usepackage{subfigure}
\usepackage{enumerate}
\usepackage{wrapfig}
\graphicspath{{./}{./Plots/}}

\newcommand{\nn}{\nonumber}
\newcommand{\beq}{\begin{equation}}
\newcommand{\eeq}{\end{equation}}
\newcommand{\bea}{\begin{eqnarray}}
\newcommand{\eea}{\end{eqnarray}}

\def\sla#1{\setbox0=\hbox{$#1$}\dimen0=\wd0
      \setbox1=\hbox{/} \dimen1=\wd1 \ifdim\dimen0>\dimen1
      \rlap{\hbox to \dimen0{\hfil/\hfil}} #1
      \else
      \rlap{\hbox to \dimen1{\hfil$#1$\hfil}}
      /   \fi}

\newcommand{\Min}{{\rm min}}
\newcommand{\Max}{{\rm max}}
\newcommand{\maos}{\rm maos}

\title{\boldmath Reducing combinatorial uncertainties:\\A new technique 
based on $M_{T2}$ variables}

\author[a]{Kiwoon Choi,}
\author[b]{Diego Guadagnoli,}
\author[c]{and Chan Beom Park}

\affiliation[a]{Department of Physics, KAIST, Daejeon 305-701, Korea}
\affiliation[b]{Laboratoire de Physique Th\'eorique, Universit\'e
  Paris-Sud, Centre d'Orsay,\\F-91405 Orsay-Cedex, France}
\affiliation[c]{Instituto de F\'isica Te\'orica UAM/CSIC,
  Nicol\'as Cabrera 13-15, Universidad Aut\'onoma de Madrid,\\
  Cantoblanco, 28049
  Madrid, Spain}

\emailAdd{kchoi@kaist.ac.kr}
\emailAdd{diego.guadagnoli@th.u-psud.fr}
\emailAdd{chanbeom.park@csic.es}

\abstract{We propose a new method to resolve combinatorial ambiguities 
in hadron collider events involving two invisible particles in the final 
state. This method is based on the kinematic variable $M_{T2}$ and on
the $M_{T2}$-assisted-on-shell reconstruction of invisible momenta,
that are reformulated as `test' variables $T_i$ of the correct combination against 
the incorrect ones. We show how the efficiency of the single $T_i$ in providing the
correct answer can be systematically improved by combining the different $T_i$ and/or 
by introducing cuts on suitable, combination-insensitive kinematic variables.
We illustrate our whole approach in the specific example of top anti-top production, 
followed by a leptonic decay of the $W$  on both sides. However, by construction,
our method is also directly applicable to many topologies of interest 
for new physics, in particular events producing a pair of undetected 
particles, that are potential dark-matter candidates. We finally emphasize 
that our method is apt to several generalizations, that we outline in the 
last sections of the paper.}


\begin{document}

\begin{flushright}
  LPT Orsay 11-77\\
  IFT-UAM/CSIC-11-64
\end{flushright}

\vskip-2.38em
\maketitle

\vspace{-0.48cm}
\section{Introduction}

Many models providing WIMP dark-matter candidates predict events at 
the Large Hadron Collider (LHC), characterized by a pair of invisible particles 
in the final state. 
This very same topology is also available in the Standard Model
(SM), e.g. in dileptonic decays of a top-quark pair: $t \bar t
\to b \ell^+ \nu \bar b \ell^- \bar \nu$. Several kinematic methods have been proposed 
to determine the particle masses in such missing energy events~\cite{Barr:2010zj}. 
On the other hand, to access information beyond the mass, e.g. the spin and/or
chirality of parent particles, it is very often highly desirable, sometimes 
necessary, to reconstruct the parent particles' full momenta event by event. 
Such reconstruction generically suffers from the combinatorial ambiguity
of correctly assigning visible particles to the given event
topology. For instance, there are two possibilities to pair the
two $b$-jets and the two charged leptons in the dileptonic decay of a
top quark pair, and three possibilities to group the jets into two dijets 
in the gluino pair decay
$\tilde{g}\tilde{g} \rightarrow jj\tilde\chi_1^0 jj \tilde\chi_1^0$
in supersymmetric models, $\tilde\chi_1^0$ denoting the LSP
neutralino. Not to mention that the possible combinations proliferate 
when the event involves multi-step cascade decays and one wishes to 
identify the order of visible particles in each decay chain, besides the 
assignment to the correct decay chain. For instance, in the event 
$\tilde\chi_2^0 \tilde\chi_2^0 \rightarrow \tilde\ell^{\pm} \ell^{\mp}
\tilde\ell^{\pm} \ell^{\mp}
\rightarrow \ell^+ \ell^- \tilde\chi_1^0 \ell^- \ell^+ \tilde\chi_1^0$, 
one has in total $2\times 2\times 2=8$ possibilities to assign the charged 
leptons to the event topology (assuming they have all the same flavor), 
and even more to correctly assign jets in the cascade decay
$\tilde{g}\tilde{g} \rightarrow j \tilde q  j \tilde q \rightarrow jj
\tilde\chi_1^0 jj \tilde\chi_1^0$ ($3\times 2\times 2=12$).

Quite generally, it is clear that minimizing combinatorial uncertainties is 
one of the first steps toward an accurate reconstruction of the full parent 
particles' momenta in missing energy events.\footnote{%
For example, see the method recently proposed in ref.~\cite{Rajaraman:2010hy}.}
Actually, this problem is already 
relevant within the SM, with no need to invoke new-physics events.
A well-known example is that of dileptonic $t\bar t$ events in the SM, 
where the reconstruction of the $t$ and $\bar{t}$ momenta is required 
in order to study spin correlations in $t \bar{t}$
\cite{Mahlon:1995zn,Stelzer:1995gc,Mahlon:1997uc,Mahlon:2010gw},
which in turn allow to probe the quantum structure of the decay
much more thoroughly than the information from the cross section
alone does. Broadly speaking, strategies aimed at fully reconstructing
the {\em momenta} of parent particles in missing-energy events are of
obvious relevance when it comes to making the most out of the LHC data.

In this paper, we aim to develop a model-independent way of reducing
combinatorial uncertainties for missing energy events at the LHC.
We propose a new method based on the combined use of the kinematic 
variable $M_{T2}$~\cite{Lester:1999tx,Barr:2003rg} and of 
the $M_{T2}$-assisted-on-shell (MAOS) reconstruction of missing 
momenta~\cite{Cho:2008tj}. All these kinematic variables are reformulated
as `test' variables $T_i$ (testing the correct pairing of final-state visible
particles against the incorrect ones) and the information from the
various $T_i$ is then used combinedly, in order to improve over the
performance of each $T_i$ separately.

We apply this strategy to the dileptonic $t\bar{t}$ decay, that, as
mentioned, represents a prototype example of pair production of a parent 
particle of interest decaying into partly-invisible daughters. Besides
its intrinsic interest, the $t \bar t$  decay offers a popular event 
topology in SM extensions with a dark-matter candidate, rendered stable 
by a conserved $Z_2$-like symmetry. 
Applying our method to dileptonic $t\bar{t}$, we find
an efficiency of determining the correct partition in the ballpark of
90\%, and that this efficiency can be systematically increased by introducing
cuts on certain partition-insensitive kinematic variables, at the price
of an actually moderate loss in event statistics.

The organization of our paper is best summarized in the table of content.
In sec. \ref{sec:testvariables} we start with an example of the problem at 
hand, and then introduce our test variables $T_i$ in terms of $M_{T2}$ and of the
MAOS algorithm, applied to the full $t \bar t$ decay or to its $WW$ subsystem.
We here examine the efficiency of each variable $T_i$ in the context of a 
parton-level analysis. In sec. \ref{sec:improvingmethod}, we address the issue
of improving over the efficiency of each single $T_i$, by devising a 
combined test, and introducing the above mentioned cut variables.
In section \ref{sec:generalization}, we discuss some generalizations
of our method, in particular its application to generic missing energy events 
producing a pair of dark matter particles in the final state. Finally, 
in sec. \ref{sec:conclusions} we conclude, providing an outlook of future work.

\section{Test variables to choose among partitions of the visible final states}
\label{sec:testvariables}

\subsection{An example} \label{sec:example}

We will state our problem of interest in the example of $t \bar{t}$
production followed by fully leptonic $W$ decays. The decay chain is
namely
\beq
\label{eq:ttbardecay}
t \bar{t} \to b W^+ ~ \bar{b} W^- \to b \ell^+ \nu ~ \bar{b} \ell^-
\bar{\nu}~.
\eeq
\begin{wrapfigure}{r}{0.5\textwidth}
    \hspace{5pt}
    \includegraphics[width=0.49\textwidth]{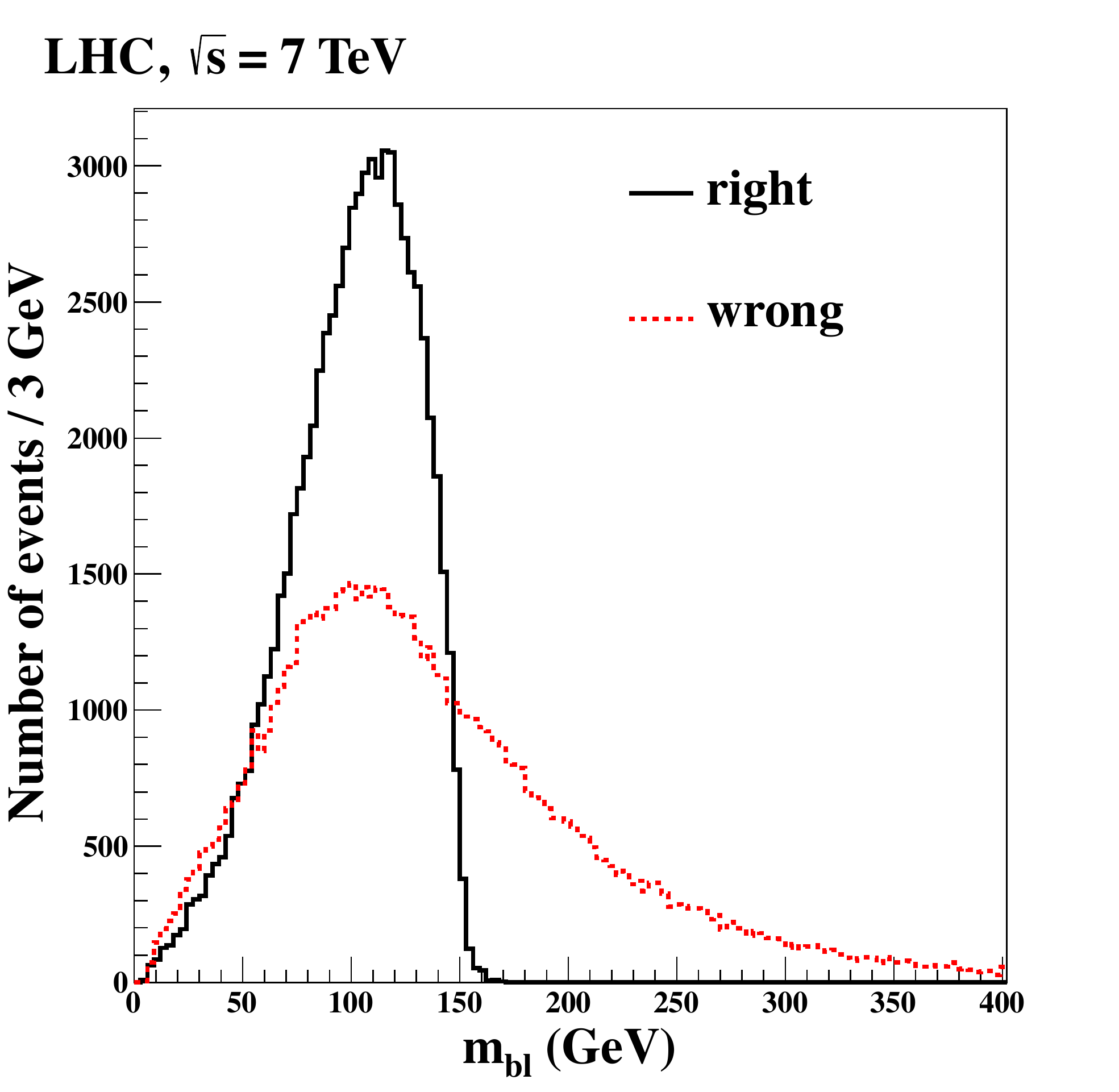}
    \caption{The $m_{b\ell}$ distribution for the $t \bar{t}$ event 
    topology (\ref{eq:ttbardecay}) with right (solid black) and wrong 
    (red dashed) partitions of the $b$ quarks and charged leptons.
    The event generation details for this and the rest
    of our plots are described in sec.~\ref{sec:eventgeneration}.}
  \label{fig:example}
\end{wrapfigure}
At the level of reconstructed objects, this process is triggered on
by, e.g., requiring at least two jets (two of which possibly
$b$-tagged), two leptons and missing energy in the final
state. Clearly, the triggered-on objects can be partitioned in two
different ways, namely
\bea
\label{eq:partitions}
\mbox{right partition}  & \equiv & \{ \ell_1 b_1 \} \& \{ \ell_2 b_2 \}~,\nn \\
\mbox{wrong partition} & \equiv & \{ \ell_1 b_2 \} \& \{ \ell_2 b_1 \}~,\nn
\eea
where the index labels the decay chain. In the rest of the text we
will refer to the different ways of grouping visible final states into
two sets as partitions.

Our aim is to construct test variables able to distinguish the right
from the wrong final-state partitions. To this end, one may exploit
suitable event variables, that namely display certain predictable
features as a function of the event kinematics -- like edges, thresholds
or peaks. The main observation to take advantage of is that these
features are present, and calculable from the measured final states,
{\em provided} the partition of the visible particles is the right one,
whereas they are partly or completely destroyed if the final-state
partition is incorrect.

The arguably simplest and best-known example to illustrate such ideas
is the invariant mass of the visible particles, $m_{V_i}$, belonging to
a given decay chain $i$. In the case of the event topology (\ref{eq:ttbardecay}),
the visible particles of each decay chain are a $b$-quark and a charged
lepton. The corresponding invariant mass distribution enjoys an upper
endpoint given by the formula
\bea
  \label{eq:mblmax}
  \left( m_{b\ell}^{\rm max} \right)^2 &=&
  m_b^2 + \frac{1}{2}\left(m_t^2 - m_W^2 - m_b^2\right)
  +\frac{1}{2}\sqrt{\left(m_t^2 - m_W^2 - m_b^2\right)^2
    -4m_W^2 m_b^2} \nn \\ 
  &\simeq & 153.2\,\mbox{GeV}~.
\eea
This endpoint feature is clearly displayed by the black solid
histogram in fig. \ref{fig:example}, representing the $m_{b\ell}$
distribution when correctly partitioning the final states. The same
feature does not need to hold in case $m_{b\ell}$ is calculated with
the incorrect final-state partition: the corresponding histogram is also
shown as a red dashed curve. 
Indicating the two possible partitions as $P_1$ and $P_2$, one can
then construct the {\em difference} between the $m_{b\ell}$ values
calculated with either partition, namely
\beq
\Delta T(P_2,\,P_1) \equiv T(P_2) - T(P_1)~,
\label{eq:Tdiff}
\eeq
where 
\beq
T(P_i) \equiv m_{b\ell}(P_i)
\label{eq:T1example}
\eeq
is the test variable for a given partition $P_i$.
From our previous considerations, one can expect that, if $P_1$
corresponds to the right partition, $P_R$, and $P_2$ to the wrong one,
$P_W$, then $\Delta T(P_W,\,P_R)$ is expected to be  {\em larger than
  zero}, at least for a subset of the possible kinematic
configurations of the visible final states.

We therefore introduce the following criterion for using $\Delta
T(P_2,\,P_1)$ as a variable testing the correct final-state partition in
the case of dileptonic $t \bar t$ decay:

\medskip
\noindent 
{\em Given a final state of $2 b$-jets $+ 2 \ell$, they can be
  partitioned in two possible ways, to be called $P_1$ and $P_2$. If
  $\Delta T(P_2,\,P_1)$, defined as eq.~(\ref{eq:Tdiff}),
  is found to be $>0$ ($<0$), then $P_1$
  ($P_2$) may be identified as the correct partition, with a calculable
  probability $p$ of misidentification.}
\medskip

In the sections to follow we will introduce several such {\em test} 
variables $T(P)$, and for each of them estimate the 
efficiency $1-p$.  We will also see that this efficiency
can be systematically improved in two directions:
\begin{itemize}
\item[\em (i)] by introducing further kinematic cuts on the event
  sample, at the price of a reduced statistics;
\item[\em (ii)] by combining the information from several $T(P)$
  in a global test variable, for example, a likelihood test with the
  $\Delta T(P_2,\,P_1)$ distributions interpreted as probability
  distributions.
\end{itemize}
Both of the above points will be addressed in the sections starting
from sec. \ref{sec:improvingmethod}.
All the numerical results of our paper rely on event simulations, whose 
details are introduced in the next section \ref{sec:eventgeneration}.
In the rest of sec. \ref{sec:testvariables} we will introduce
all the test variables and discuss variations around their definitions
as well as their efficiencies.

\subsection{Event generation and selection}\label{sec:eventgeneration}

We generated 100,000 parton-level events of top quark pair production
followed by dileptonic final states using {\tt MadGraph 5}, with
proton-proton collisions at $\sqrt{s} = 7$ TeV~\cite{Alwall:2011uj}.
We here do not assume any contribution from physics beyond the
SM. The SM predicts $\sigma_{t\bar{t}} \sim 162$ pb
at next-to-next-to-leading-order~\cite{Aliev:2010zk}.\footnote{The cross
  section was calculated with $m_t = 173$ GeV and CTEQ6.6 parton
  distributions~\cite{Nadolsky:2008zw}.}
Hence the generated events correspond to $\sim 13$ fb$^{-1}$. 
The dilepton mode gives rise to final states with two $b$-tagged jets,
two leptons, missing transverse energy. We consider the leptonic
decays of the $W$ bosons to an electron or muon.\footnote{There are
  small contributions from the leptonic decays of the tau, for
  example, $\tau^- \to \mu^- \bar{\nu}_\mu \nu_\tau$, but here we
  neglect them for the sake of discussion.}

We apply event selection cuts on the simulated event sample following
the 0.7 fb$^{-1}$ ATLAS dileptonic top analysis~\cite{ATLAS-CONF-2011-100}. 
We namely require
\begin{itemize}
  \item exactly two oppositely-charged leptons ($ee$, $\mu\mu$,
    $e\mu$) with $p_T > 25$ GeV and $|\eta|<2.5$,
  \item at least two jets with $p_T > 25$ GeV and $|\eta|<2.5$,
  \item dilepton invariant mass, $m_{\ell \ell} > 15$ GeV,
  \item that events in the same-flavor lepton channels must satisfy
    the missing transverse energy condition ${\sla E}_T > 60$ GeV and
    $|m_{\ell\ell} - m_Z| > 10$ GeV,
  \item that events in the different-flavor channel satisfy $H_T >
    130$ GeV. The event variable $H_T$ is defined as the scalar sum of
    the transverse momenta of the two leptons and all selected
    jets. In analogy with \cite{ATLAS-CONF-2011-100} we instead do not
    require $\sla E_T$ or $m_{\ell \ell}$ cuts in this channel.
\end{itemize}
We found that about 37\% of the total simulated events pass these
basic selection cuts. The motivation for including the mentioned cuts
is to have an event sample that resembles as much as possible the
actual experimental event samples in dileptonic $t \bar t$
production. As such, it will be our reference event sample throughout
our analysis. We leave aside for the moment the important issues of:
inclusion of hadronization, QCD radiation, and modeling of 
detector effects, that we will address in a forthcoming, more in-depth
study~\cite{futureWork}.

\subsection{Test variables: definition and discussion}\label{sec:Tidef}

\subsubsection[Invariant mass of visible final states: $T_1$ variable]%
{\boldmath Invariant mass of visible final states: $T_1$ variable} \label{sec:T1}
\begin{wrapfigure}{b}{0.5\textwidth}
  \includegraphics[width=0.49\textwidth]{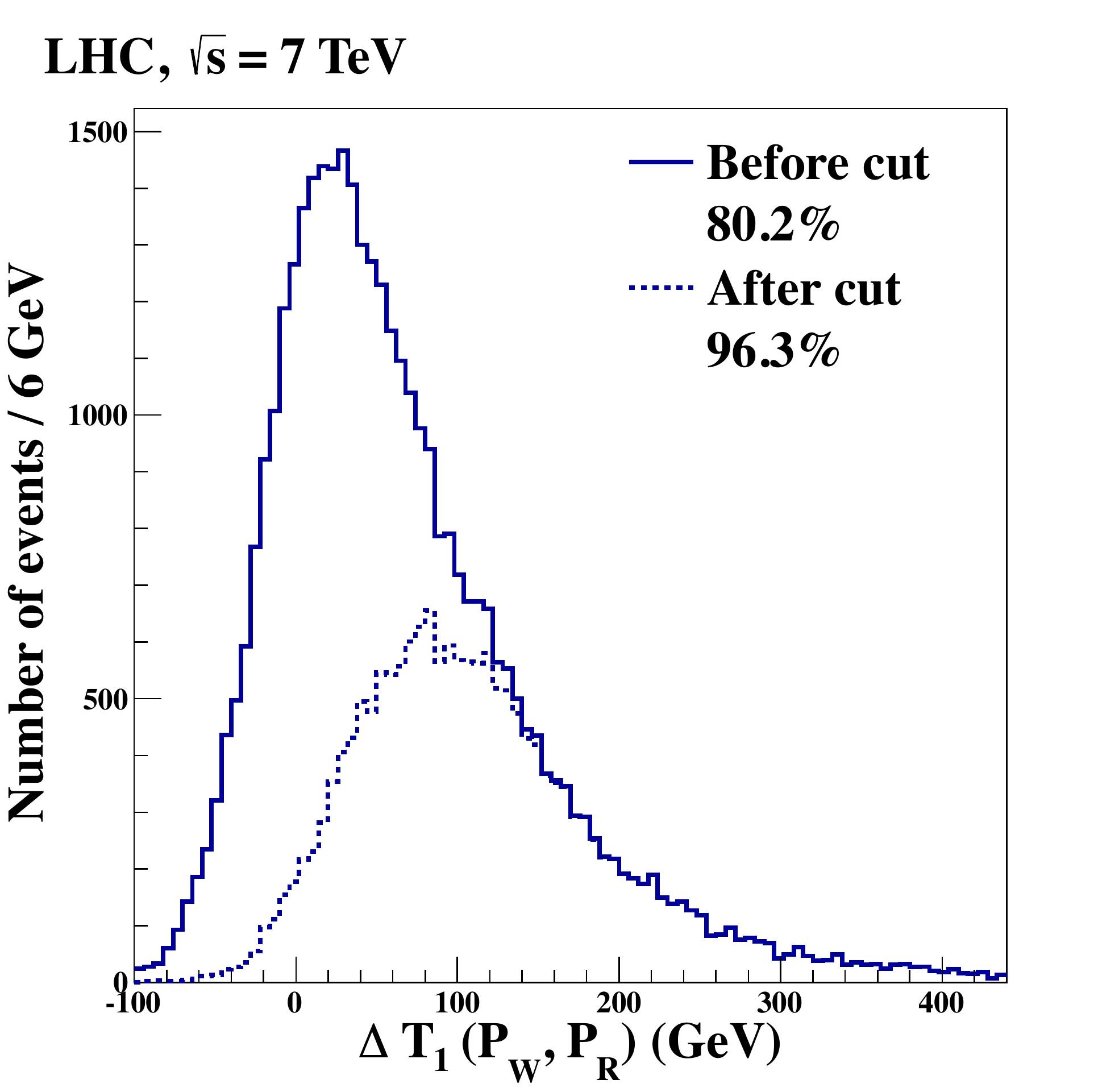}
  \caption{Distribution for the variable $\Delta T_1(P_W,\,P_R)$ (see defs. in
  eqs. (\ref{eq:T1def}) and (\ref{eq:Tdiff})) with $P_W$ and $P_R$ denoting the wrong 
  and the right partition respectively. The legend in the plot indicates 
  the percentage of correctly partitioned events, before and after 
  inclusion of the cut $M_T^{t\bar{t}}(0)>400$ GeV, to be described in sec.~\ref{sec:cut}.}
  \label{fig:mbl}
\end{wrapfigure}

Our first test variable is constructed from $m_{b\ell}$, mentioned in the example of 
sec.~\ref{sec:example}. In particular, it could be defined directly as in 
eq.~(\ref{eq:T1example}). However, for each event one obtains two $m_{b\ell}$ values, 
$m_{b\ell}^{(i)}$ $(i = 1,\,2)$, corresponding to the two decay chains. Combining the two
$m_{b\ell}^{(i)}$ values in different ways gives rises to alternative definitions of the 
test variable, not all of which yield the same efficiency. We found the combination
$\Max[m_{b\ell}^{(1)},\,m_{b\ell}^{(2)}]$ to perform best:
\begin{eqnarray}
\label{eq:T1efficiency}
  &&\Max\left[m_{b\ell}^{(1)},\,m_{b\ell}^{(2)}\right]
  (P_W) > \nonumber\\
  &&\max\left[m_{b\ell}^{(1)},\,m_{b\ell}^{(2)}\right]
  (P_R) : 80.2\%~.
\end{eqnarray}
We correspondingly define our first test variable as follows:
\beq
\label{eq:T1def}
T_1(P_i) \equiv \Max\left[m_{b\ell}^{(1)},\,m_{b\ell}^{(2)}\right] (P_i)~,
\eeq
for a given partition $P_i$. The $\Delta T_1(P_W,\,P_R)$ distribution is shown 
in fig.~\ref{fig:mbl}.

The efficiency figure in eq. (\ref{eq:T1efficiency}), as well as in the solid histogram
in fig.~\ref{fig:mbl}, refers to the events that pass the basic selection cuts described
in sec.~\ref{sec:eventgeneration}, and actually includes one further consideration: if one 
of the partitions results in $m_{b\ell} > m_{b\ell}^{\rm max}$, then the right partition 
can be clearly selected because $m_{b\ell}(P_R) \leq m_{b\ell}^{\rm max}$.

\subsubsection[$M_{T2}$ of the whole decay chain: $T_2$ variable]%
{\boldmath $M_{T2}$ of the whole decay chain: $T_2$ variable}\label{sec:T2}

Another event variable that may be used to construct a test variable
of correct partitioning is $M_{T2}$, the so-called ``stransverse
mass'', which is especially suited to topologies with pair-produced
particles, each decaying into partly invisible final
states~\cite{Lester:1999tx,Barr:2003rg}. The general event topology
that $M_{T2}$ is suited for is the following:
\beq
\label{eq:eventTopology}
Y \bar Y ~\to~ V_1(p_1) \chi(k_1) + V_2(p_2) \bar \chi(k_2)~,
\eeq
where each of the $V_i$ denotes a (set of) visible state(s), whose
total four-momentum is, in principle, entirely reconstructible, and
$\chi$, $\bar \chi$ denote invisible particles with identical mass.
The decay mode (\ref{eq:ttbardecay}) is an example of this very event topology: 
$V_i$ should be identified with a $b~\ell$ pair (for each decay chain) and
$\chi, \bar \chi$ with the neutrinos.

For the event topology (\ref{eq:eventTopology}), the $M_{T2}$ variable
is defined as
\bea
  \label{eq:MT2def}
  M_{T2}(\tilde m_\chi) \equiv \min_{ \tilde{\bf k}_{{\bf 1}T} + \tilde{\bf k}_{{\bf 2}T} = 
  {\bf{\sla p}}_T} \left[ \max \left\{ 
  M_T^{(1)} \left({\bf p_1}_T,\, \tilde{\bf k}_{{\bf 1}T},\, p_1^2,\,\tilde{m}_\chi\right),\,
  M_T^{(2)} \left({\bf p_2}_T,\, \tilde{\bf k}_{{\bf 2}T},\, p_2^2,\,\tilde{m}_\chi\right) 
  \right\} \right]~,\nn \\
\eea
where $M_T^{(i)}$ is the transverse mass
\cite{Smith:1983aa,Barger:1983wf} of the decay chain
$i\,(=1 \mbox{ or } 2)$, namely
\bea
  \left(M_T^{(i)}\right)^2 =
  p_i^2 + \tilde{m}_\chi^2 + 2 \left(
  \sqrt{|{\bf p_i}_T|^2 + p_i^2}\sqrt{|\tilde{\bf k_i}_T|^2 +
  \tilde{m}_\chi^2} - {\bf p_i}_T\cdot\tilde{\bf k_i}_T\right)~.
  \label{eq:MTi}
\eea
A few comments may help get an intuitive picture of the above
definitions. First, it should be kept in mind that what is measured in the
event topology of (\ref{eq:eventTopology}) are the visible
particles' momenta, $p_i$, and the total momentum imbalance in the
transverse direction, ${\bf \sla p}_T$. 
The definition in eq.~(\ref{eq:MT2def}) correspondingly marginalizes
over the transverse momenta ${\bf k}_{{\bf 1,2}T}$ of $\chi$, $\bar\chi$, 
that are not separately measured, by taking the minimum over all the 
{\em trial} momentum configurations (denoted with a tilde), whose sum
equals the measured ${\bf \sla p}_T$. 

Second, the invisible mass $m_\chi$ is, likewise, unmeasured, hence $M_{T2}$
is a function of a trial value for this mass, $\tilde m_\chi$. 
In the case of the decay (\ref{eq:ttbardecay}) $m_\chi$
is of course known: $m_\chi = m_\nu \approx 0$. However, the topology in 
eq. (\ref{eq:eventTopology}) applies also to several new-physics scenarios,
and in this case the physical mass $m_\chi$ of the pair-produced $\chi$ is 
in general unknown.
An interesting feature of $M_{T2}$ is the kink structure of 
$M_{T2}^{\rm max} (\tilde m_\chi)$ at $\tilde m_\chi = m_\chi$, provided 
the $p_i^2$ value spans a certain range in each decay 
chain~\cite{Cho:2007qv,Cho:2007dh}. An alternative possibility for the 
kink structure of $M_{T2}^{\rm max}(\tilde m_\chi)$ to appear is to have
the $Y\bar{Y}$ system boosted by a sizable amount of upstream transverse
momentum~\cite{Gripaios:2007is,Barr:2007hy}.
The $M_{T2}$-kink method makes it possible to measure $m_\chi$ and
$m_Y$ {\em simultaneously}, and even if the decay chain is not long enough
to constrain all the unknowns in the event.

The property of $M_{T2}$ most relevant to our discussion is that, when the
input trial mass $\tilde{m}_\chi$ equals the true mass value $m_\chi$,
the upper endpoint of the $M_{T2}$ distribution corresponds to the
parent particle mass $m_Y$. We want to use this property to 
distinguish the correct against the incorrect partition of the
final-state visible particles. We note explicitly that, in dileptonic
$t\bar{t}$, one can construct two kinds of $M_{T2}$:
$M_{T2}^{t\bar{t}}$ for the full $t\bar{t}$ system and $M_{T2}^{WW}$
for the $WW$ subsystem~\cite{Nojiri:2008vq,Burns:2008va,Choi:2010dk,Choi:2010dw},
the visible particles being respectively $b \ell^+ \bar{b} 
\ell^-$ and $\ell^+ \ell^-$. Since only one partition is possible in 
$M_{T2}^{WW}$, this variable cannot be used directly as a test
variable. It can however be used to construct other test variables, to
be introduced in the next section. 
In the rest of the present discussion we will therefore specialize to
$M_{T2}^{t\bar{t}}$.

\begin{figure}[t!]
  \begin{center}
  \includegraphics[width=0.49\textwidth]{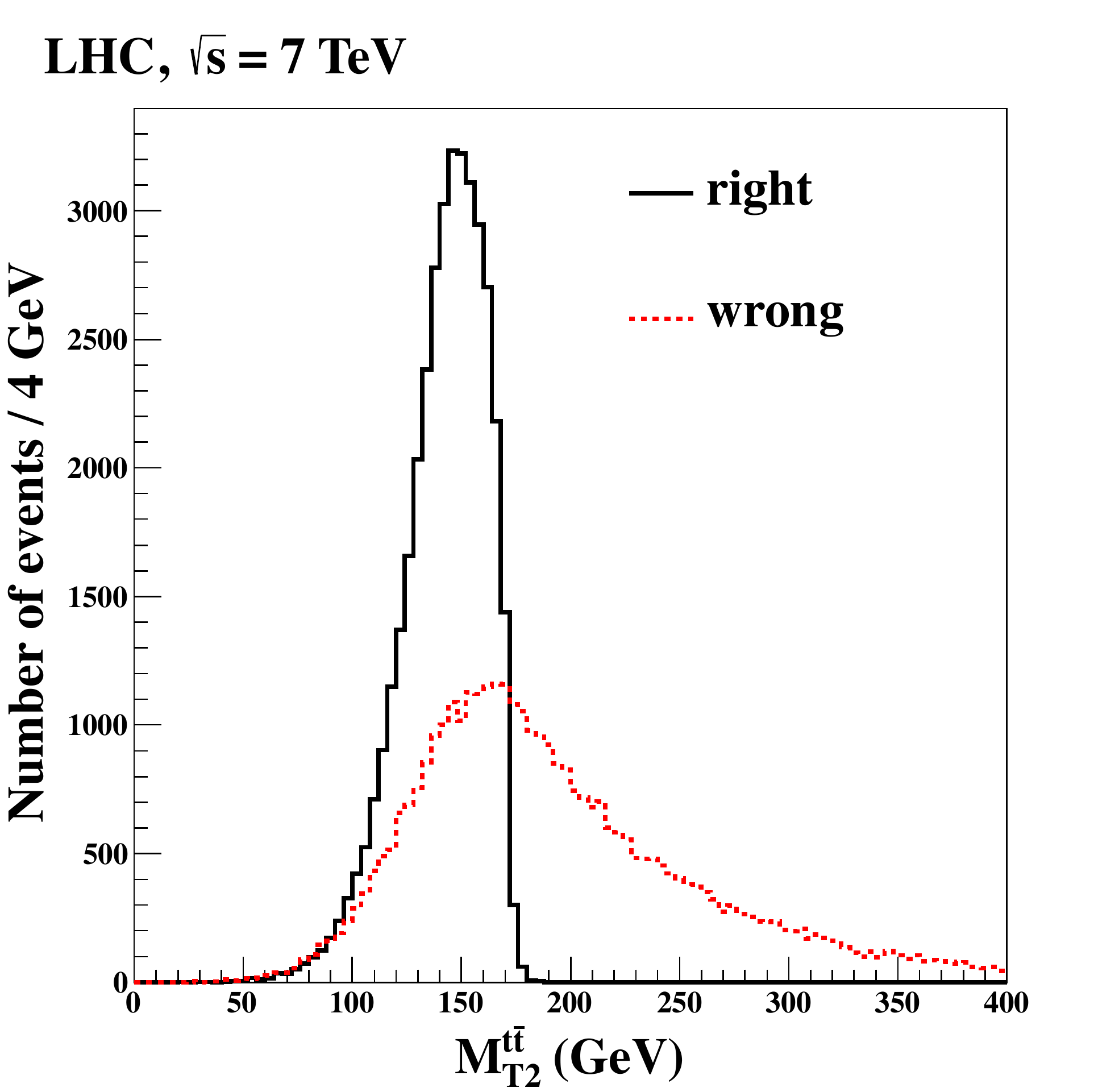}
  \includegraphics[width=0.49\textwidth]{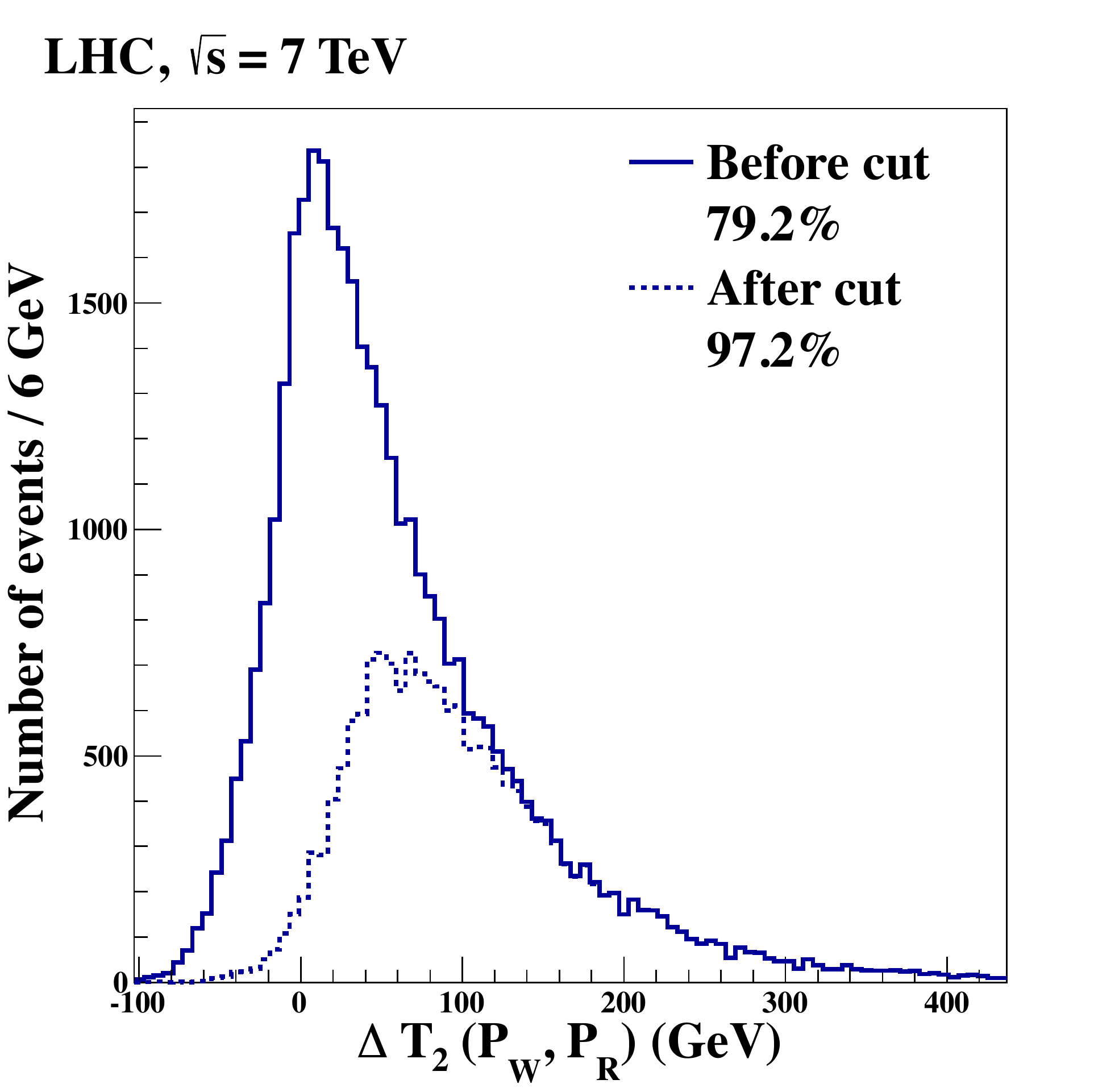}
  \end{center}
  \caption{The distributions of $T_2 \equiv M_{T2}^{t\bar{t}}$
      (left panel) and $\Delta T_2(P_W,\,P_R)$ (right panel), with $P_W$ and 
      $P_R$ the wrong and right partitions respectively. The legend in the second
      panel indicates the percentage of correctly partitioned events, 
      before and after inclusion of the cut $M_T^{t\bar{t}}(0)>400$ GeV, to be 
      described in sec.~\ref{sec:cut}.}
  \label{fig:T2}
\end{figure}

As already mentioned, the $M_{T2}^{t\bar{t}}$ distribution is bounded
from above by the top quark mass $m_t$ if the partition is the right
one, whereas it is not if the partition is the wrong one.
This fact is shown in the left panel of fig. \ref{fig:T2}. One can therefore 
define the variable
\beq
T_2(P_i) \equiv M_{T2}^{t\bar{t}} (P_i)~,
\label{eq:T2def}
\eeq
where we recall that $P_{i}$ ($i=1,\,2$) denotes the two possible
partitions of the $2b + 2\ell$ visible final states.

In the right panel of fig.~\ref{fig:T2}, we show the
distribution of $\Delta T_2(P_W,\,P_R)$ with the right ($P_R$) and
wrong ($P_W$) partitions.
The figure shows that the relation $\Delta T_2(P_W,\,P_R)>0$ holds
for about 80\% of the events that passed just the basic selection cuts 
described in sec. \ref{sec:eventgeneration}, and that this percentage reaches 
as much as 97\% after the $M_T^{t\bar{t}}(0)$ cut to be described in 
sec. \ref{sec:cut}.

To conclude this section, we mention that the $M_{T2}$ method of finding the right partition 
has been proven to be useful for mass measurements and/or event reconstruction in
refs.~\cite{Barr:2007hy,Lester:2007fq,Cho:2008cu,Alwall:2009zu,Aaltonen:2009rm,
Konar:2009qr,Berger:2010fy,Nojiri:2010mk,Zhang:2010kr,Berger:2011ua,Park:2011uz,
Murayama:2011hj}.

\subsubsection[MAOS-reconstructed $m_t$ and $m_W$ mass: $T_{3,4}$ variables]{\boldmath 
MAOS-reconstructed $m_t$ and $m_W$ mass: $T_{3,4}$ variables}\label{sec:T34}

A further set of variables to test the correct partition can be
constructed from the observation that the sum of final-state momenta
in either decay chain must reconstruct, up to width effects, the
parent-particle's invariant mass. The main obstacle to this, in the
case of the event topology (\ref{eq:eventTopology}), is the
fact that the invisible momenta of the two decay chains, $k_1$ and
$k_2$, are not measured separately, but only their sum is, and only 
in the transverse plane.

Systematic techniques exist however, enabling to obtain a {\em best guess} 
of the invisible momenta $k_1$ and $k_2$.\footnote{%
For the dileptonic $t\bar{t}$ decay, one of the presently most popular
techniques is the neutrino weighting method, which has been used for
the top quark mass measurement~\cite{Abbott:1997fv,Abazov:2009eq} 
and for studying spin correlations in $t\bar{t}$
production~\cite{Abazov:2011qu}.
}
Among these techniques is the so-called $M_{T2}$-assisted on-shell
(MAOS) reconstruction of invisible momenta~\cite{Cho:2008tj}. The only
assumption of the MAOS method is that the event topology should be of
the kind of eq. (\ref{eq:eventTopology}). Then the transverse
components of the invisible momenta, ${\bf k}_{{\bf 1}T}$, ${\bf
  k}_{{\bf 2}T}$, can be estimated event by event to be the location
of the minimum of the $M_{T2}$ variable constructed for the
event. Recall in fact that, from eq. (\ref{eq:MT2def}), $M_{T2}$ is
obtained from a minimization over all possible $\tilde {\bf k}_{{\bf 1,2}T}$
configurations subject to the constraint $\tilde {\bf k}_{{\bf 1}T} +
\tilde {\bf k}_{{\bf 2}T} = {\bf {\sla p}}_T$.

Once ${\bf k}_{{\bf 1}T}$ and ${\bf k}_{{\bf 2}T}$ are estimated by $M_{T2}$,
the longitudinal components of the invisible momenta may be determined 
from the on-shell relations 
\bea
\label{eq:MAOSonshellrelations}
(p_i + k_i^{\maos})^2 = m_Y^2~,\quad
(k_i^{\maos})^2 = m_\chi^2
\eea
where, as usual, $i$ labels the decay chain, $m_Y$ the parent particle
mass, and $m_\chi$ the invisible particle mass. 
Eqs. (\ref{eq:MAOSonshellrelations}) amount to a quadratic
equation in the longitudinal components of $k_i$, hence one has in
general two solutions for each decay chain, $\tilde{k}_{{\bf i}L}^{\pm}$.
Modulo this two-fold ambiguity, and assuming $m_{\chi}$ to be known (in
our case it is $m_\nu \approx 0$), the MAOS reconstruction thus leads
to a estimate of the {\em full} invisible momenta $k_i$. Specifically,
$k_i^{\maos}$ has a roughly Gaussian distribution around $k_i^{\rm
  true}$ and a moderate spread, that can be systematically improved by
requiring event cuts~\cite{Cho:2008tj}.

\begin{figure}[t!]
  \begin{center}
    \includegraphics[width=0.49\textwidth]{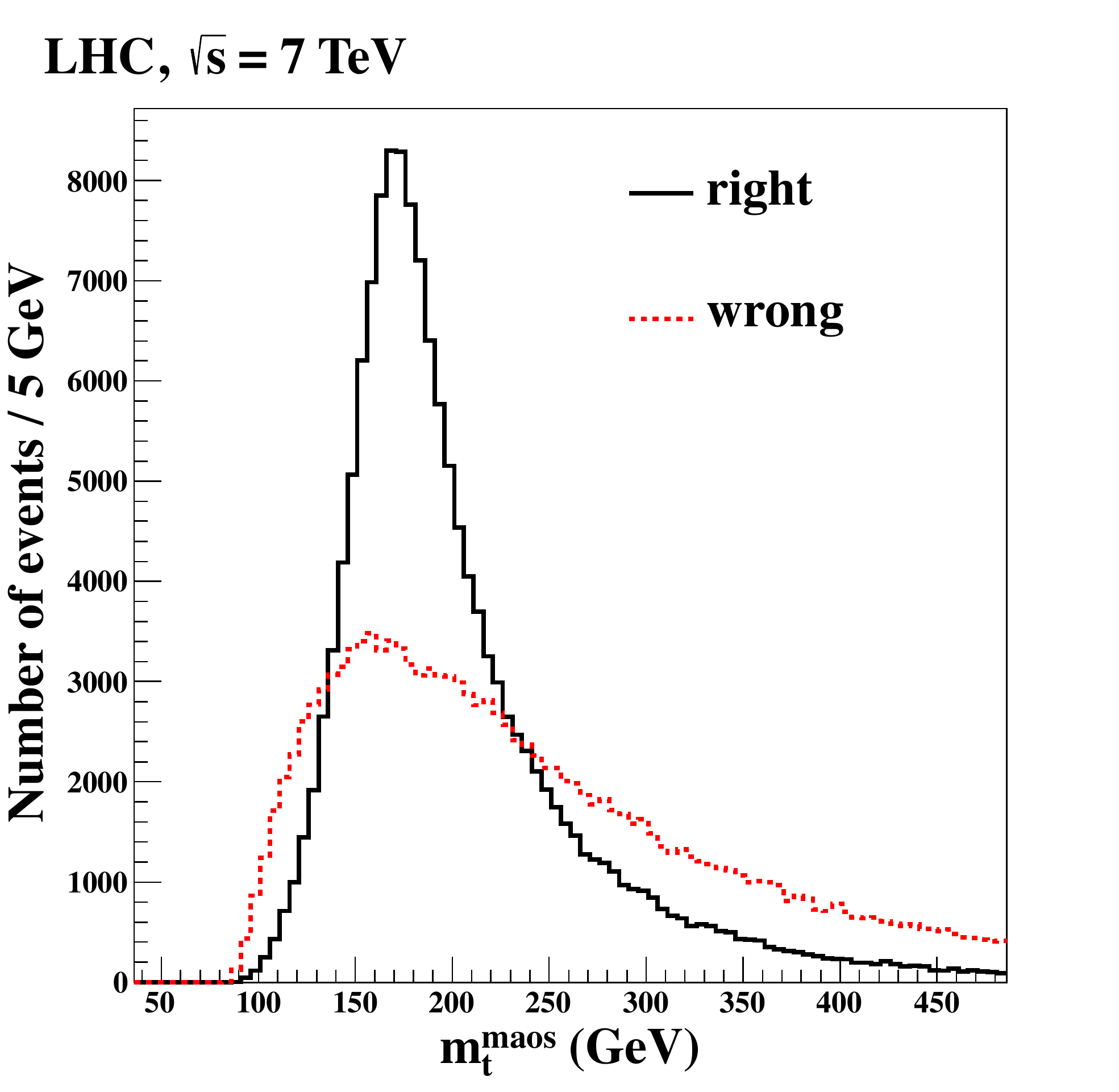}
    \includegraphics[width=0.49\textwidth]{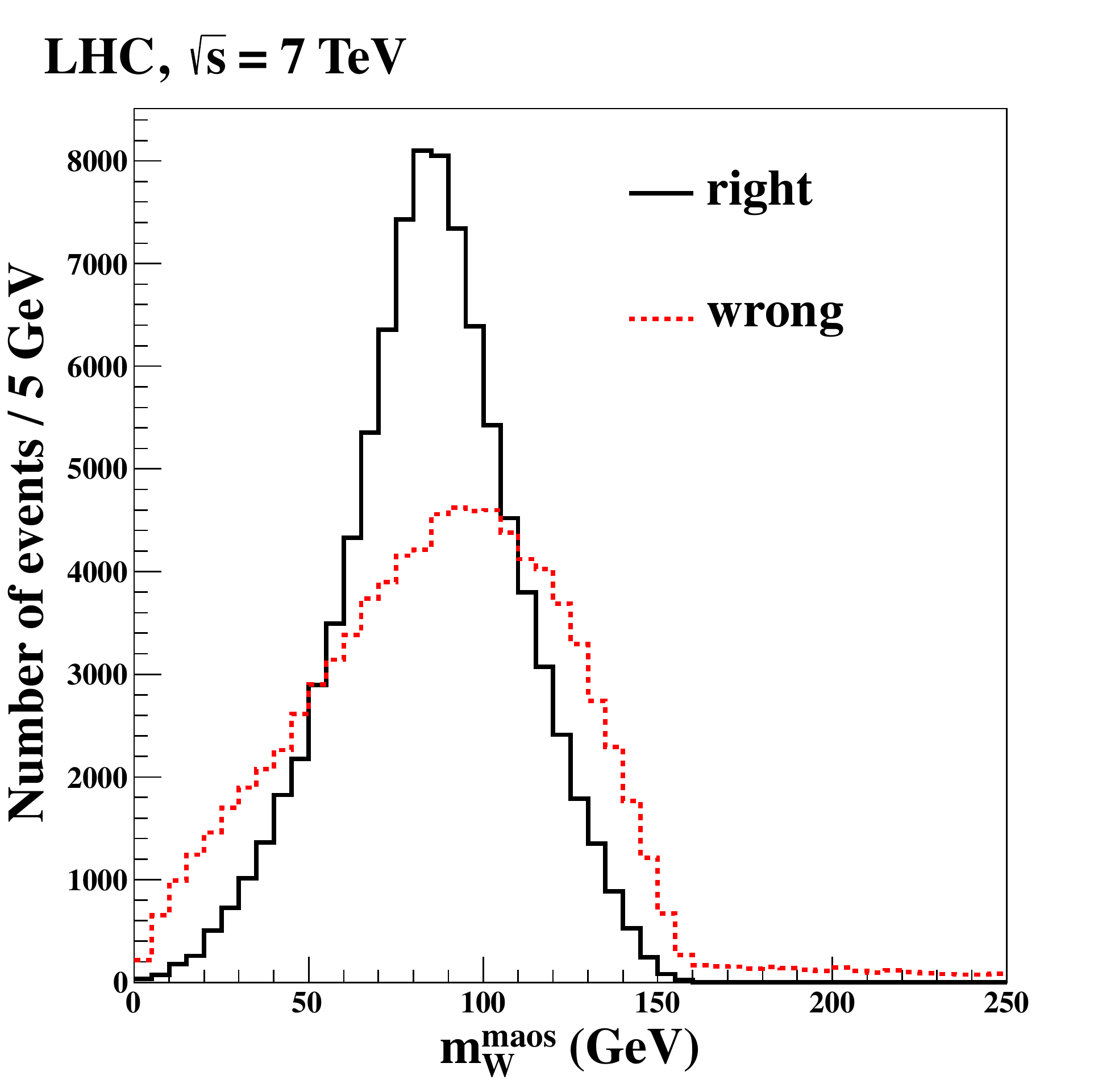}
  \caption{
    The $m_t^{\rm maos}$ (left panel) and $m_W^{\rm maos}$ (right panel) 
    distributions for the right and for the wrong final-state partition. 
    The distributions include only events with real MAOS solutions for 
    all partitions (see text for details).}
  \label{fig:maosTop}
  \end{center}
\end{figure}

In our case, the whole argument in the previous paragraph can be
applied to the full $t \bar t$ decay, or to the $W W$ subsystem,
because both share the same, mentioned topology
(\ref{eq:eventTopology}). In particular, from the MAOS momenta of the
$WW$ system, $k^{{\rm maos}\mbox{-}WW}_{1,2}$, one can construct the
invariant mass of the top quark as
\beq
  \label{eq:mtmaos}
  \left(m_{t,i}^{\maos}\right)^2
  \equiv \left(p_{b,i} + p_{\ell,i} + k_i^{{\rm maos}\mbox{-}WW} \right)^2~,
\eeq
whereas from the MAOS momenta of the $t\bar{t}$ system, $k_i^{{\maos}
  \mbox{-}t\bar{t}}$, one can construct the $W$ boson invariant mass as
\beq
  \label{eq:mWmaos}
  \left(m_{W,i}^{\maos}\right)^2
  \equiv \left(p_{\ell,i} + k_i^{{\maos}\mbox{-}t\bar{t}}\right)^2~.
\eeq
It is clear that, if the right partition is selected and in the limit where
the MAOS momenta equal the true momenta in each decay chain,
eqs. (\ref{eq:mtmaos}) and (\ref{eq:mWmaos}) should equal respectively
the true $m_t$ and $m_W$ values, up to finite decay widths. In
practice, the MAOS momenta differ in general from the true
momenta. However, the MAOS invariant mass distribution exhibits a peak
structure around the true parent particle mass. Hence --- to the
extent that the spread around the peak value is not too large --- the
above eqs.~(\ref{eq:mtmaos}) and (\ref{eq:mWmaos}) are still
useful for our purpose, which is, we recall, not to measure the $t$ or
$W$ mass, but to single out the correct final-state partition. We show
in fig.~\ref{fig:maosTop} the $m_t^{\rm maos}$ (left panel) and the
$m_W^{\maos}$ (right panel) distributions calculated using the right
partition (black solid histogram) and the wrong partition (red dashed
histogram). 

The argument below eqs. (\ref{eq:mtmaos})-(\ref{eq:mWmaos}) may be
used to construct a test variable of correct partition by noting that
\beq
\label{eq:mtmaosInequality}
|m_{t,i}^{\rm maos} - m_t| (P_W) > |m_{t,i}^{\rm maos} - m_t|(P_R)~,
\eeq
and a similar inequality holds also for $m_W^{\rm maos}$. In reality,
the notation in relation (\ref{eq:mtmaosInequality}) is
incomplete, because of the mentioned discrete ambiguity in the
determination of the longitudinal components of the invisible
momenta. This is a common problem for methods attempting to reconstruct
invisible momenta, and based on polynomial on-shell equations of degree
higher than one. One cannot distinguish the solutions among themselves,
even though only one solution corresponds to the true one and the
others are redundant.

We construct our test variable treating these solutions in a symmetric
way. We first define a quantity
\beq
  \Delta_t m_t^{\rm maos}(\alpha) \equiv m_t^{\rm
    maos}(p_b,\,p_l,\,k^\alpha) - m_t
  \quad(\alpha = +,\,-)~,
  \label{eq:deltaMaosTop}
\eeq
where we omitted the chain label, and $\alpha$ denotes the two MAOS
solutions for the invisible momentum $k^{\rm maos}$ in that decay
chain. 
Thence we found that
\beq
  \label{eq:deltamtInequality}
  \sum_{i=1,2;\, \alpha=+,-} |\Delta_t m_{t,i}^{\rm maos}(\alpha)|(P_W) 
  > \sum_{i=1,2;\, \alpha=+,-} |\Delta_t m_{t,i}^{\rm maos}(\alpha)|(P_R) :
  84.2\%~,
\eeq
namely that the sum over chains and over $k^{\pm}$ solutions of the
quantity defined in (\ref{eq:deltaMaosTop}) is smaller when
calculated on the right partition than on the wrong partition in as
much as 84\% of the events.\footnote{%
  We mention that we have tried a number of alternative combinations
  besides $\sum_{i=1,2;\, \alpha=+,-} |\cdot|$. In particular
  $\prod_{\alpha,\beta = +,-}|\cdot|$, $\sum_{i=1,2;\, \alpha=+,-}
  |\cdot|^2$, $\Min_{\alpha,\beta = +,-} |\cdot|$ and
  $\Max_{\alpha,\beta = +,-} |\cdot|$. The combination proposed in
  eq.~(\ref{eq:deltamtInequality}) is the one found to have the best
  efficiency.}
Consistently with the discussion in sec. \ref{sec:example}, we then
define our test variable $T_3$ as follows:
\beq
\label{eq:T3def}
T_3(P_k) \equiv
\sum_{i=1,2;\, \alpha=+,-} |\Delta_t m_{t,i}^{\rm maos}(\alpha)|(P_k)~.
\eeq

The whole line of reasoning below eq.~(\ref{eq:mtmaosInequality}) can
be applied to the $WW$ subsystem as well. Similarly as $m_t^{\maos}$ 
in (\ref{eq:deltaMaosTop}) one constructs the difference between 
$m_W^{\rm maos}$ and $m_W$ and symmetrizes between decay chains and 
$k^\pm$ solutions. We obtained
\beq
  \label{eq:deltamWInequality}
  \sum_{i=1,2;\, \alpha=+,-} 
  |\Delta_W m_{W,i}^{\rm maos}(\alpha)|(P_W) >
  \sum_{i=1,2;\, \alpha=+,-} 
  |\Delta_W m_{W,i}^{\rm maos}(\alpha)|(P_R) :
  86.8\%~,
\eeq
and we define our test variable $T_4$ as 
\beq
\label{eq:T4def}
T_4(P_k) \equiv
\sum_{i=1,2;\, \alpha=+,-} |\Delta_W m_{W,i}^{\rm maos}(\alpha)|(P_k)~.
\eeq

\begin{figure}[t!]
  \begin{center}
    \includegraphics[width=0.49\textwidth]{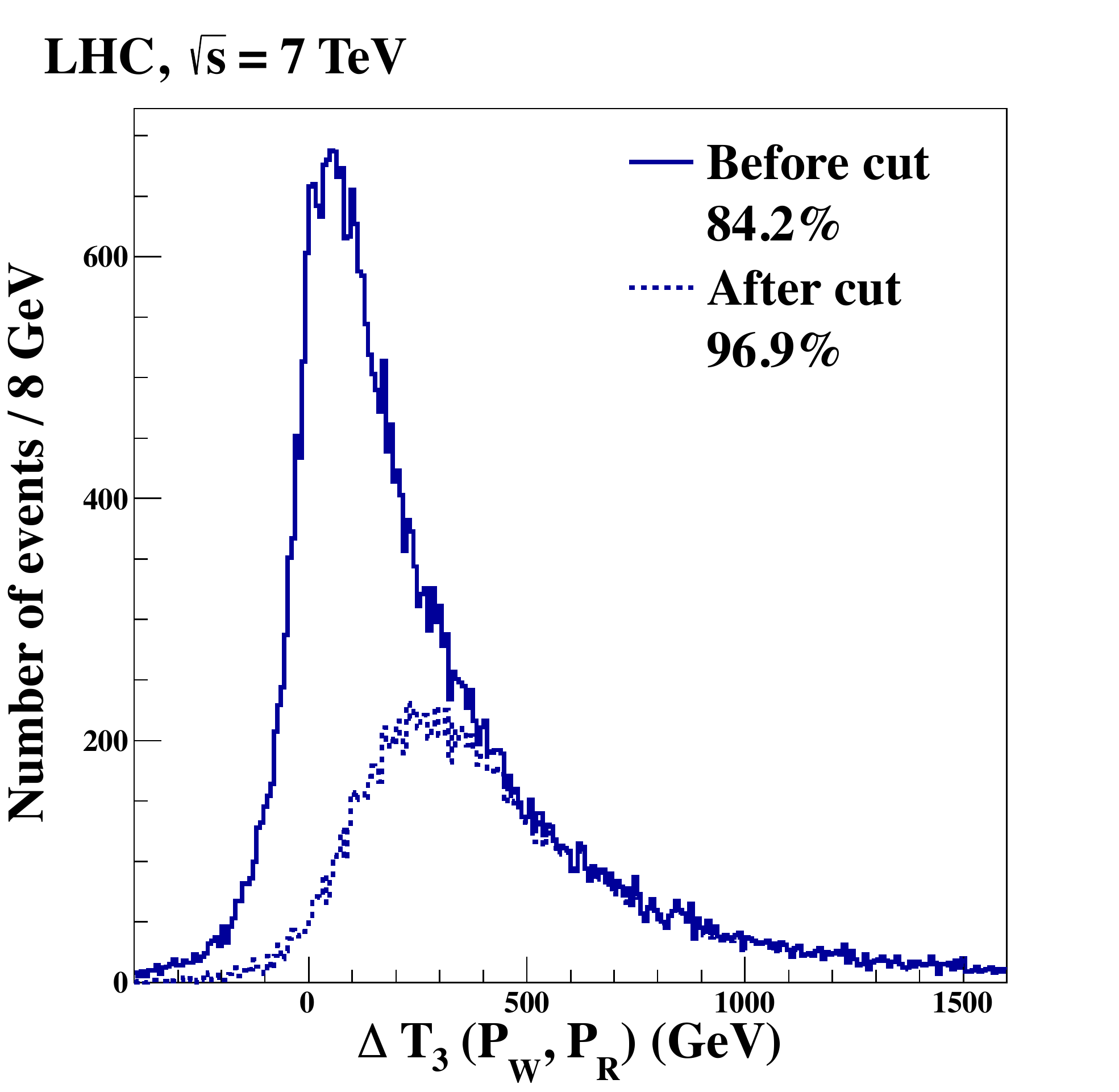}
    \includegraphics[width=0.49\textwidth]{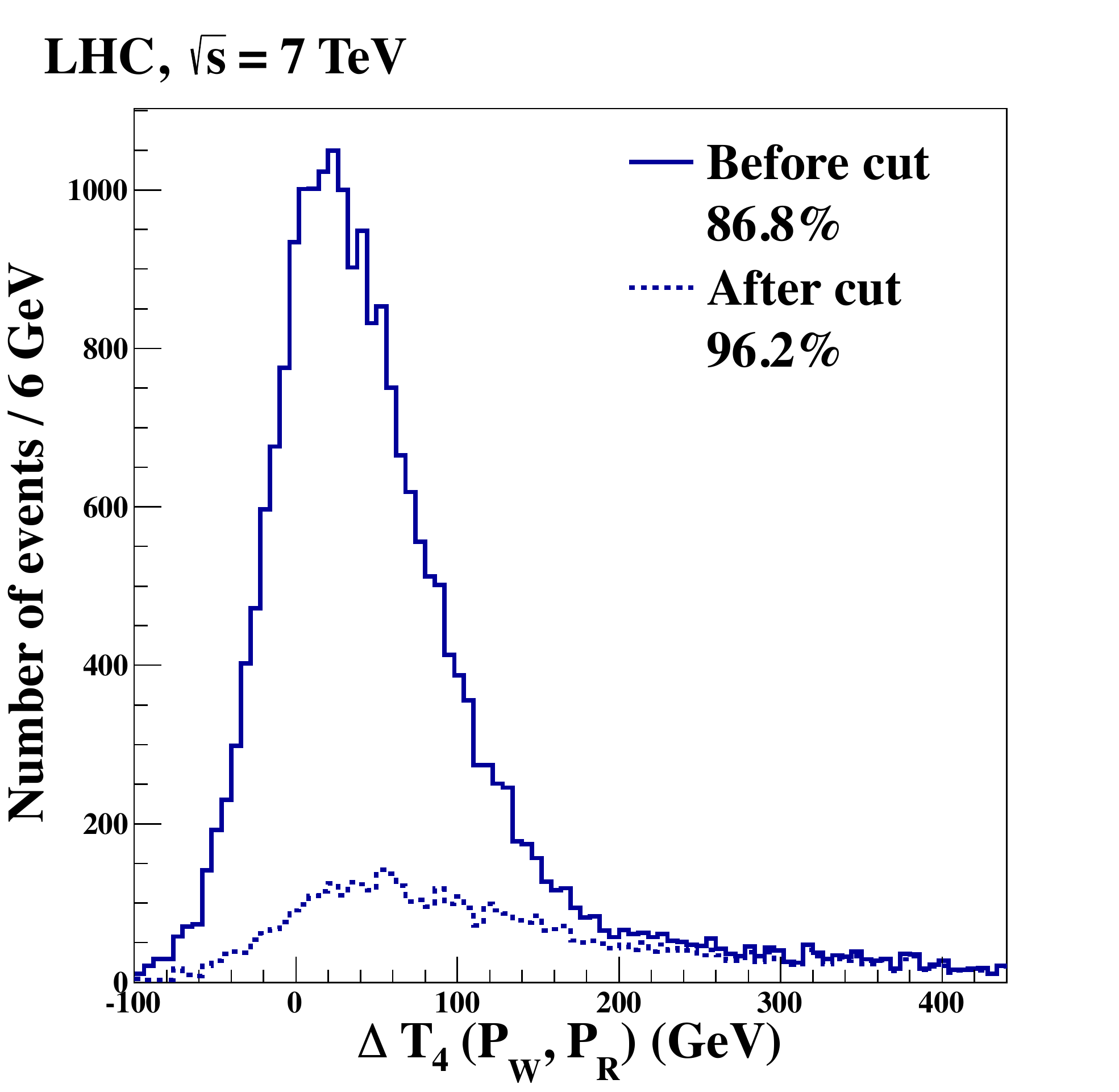}
    \caption{
        The distributions of $\Delta T_3(P_W,\,P_R)$ (left panel) and $\Delta
        T_4(P_W,\,P_R)$ (right panel), with $P_W$ and $P_R$ denoting the wrong
        and the right partition. The legend indicates the percentage of correctly 
        partitioned events, before and after inclusion of 
        the cut $M_T^{t\bar{t}}(0)>400$ GeV, to be described in sec.~\ref{sec:cut}.
        The efficiency figures in the right panel take into account the 
        complex-MAOS-solution criterion, whereas histograms report only the
        events free of complex solutions. See text for full details.
        }
    \label{fig:DeltaT3T4}
  \end{center}
\end{figure}

An important difference between the $m_t^{\rm maos}$-based method and
the $m_W^{\rm maos}$-based one should be emphasized. Within the
$m_t^{\rm maos}$ method, the choice of final-state partition enters only
when constructing the top invariant mass (see eq. (\ref{eq:mtmaos})),
whereas only one partition is possible in the estimate of 
$k_i^{{\maos}\mbox{-}WW}$. Conversely, within the $m_W^{\rm maos}$ method,
the choice of final-state partition does come into play in the
calculation of $k_i^{{\maos}\mbox{-}t \bar t}$ (see eq. (\ref{eq:mWmaos})),
namely when calculating $M_{T2}$ of the $t \bar t$ system. The
important point is that {\em wrong} partitions sometimes result in 
{\em complex} solutions for (the longitudinal components of) 
$k_i^{{\maos}\mbox{-}t \bar t}$ much more often than {\em right}
partitions do,\footnote{%
Right partitions can also yield complex solutions, but only because of
occasional failure of the numerical minimization in the $M_{T2}$
calculation, that we find to be the case only for a very small subset
of all events (about $0.6\%$). We used a modified version of the 
{\tt Mt2::Basic\_Mt2\_332\_Calculator} algorithm of the 
Oxbridge MT2 / Stransverse Mass Library~\cite{lester:mt2code}.} 
and this can be used as a further criterion for identifying the right
partition. From our simulation, we found that:
\begin{itemize}

\item[(a)] events with at least one complex solution (in any of the partitions) 
occur 38.5\% of the time;

\item[(b)] most importantly, events where a complex solution appears only in the
wrong partition occur 37.9\% of the time -- very close to the percentage in point (a).%
\footnote{Besides, we found: 
(i) events where a complex solution appears in both the right and the wrong partition 
(0.4\% of the total events). Of course for this kind of events the information from 
$m_W^{\rm maos}$ is unusable;
(ii) events where a complex solution appears only in the right partition (0.2\% of the 
total events). Events (i) and (ii) explain the difference between the percentages in 
points (a) and (b).
}

\end{itemize}

\noindent From the above points, it is clear that the $m_W^{\maos}$ method can be augmented 
by the following requirement:

\medskip
\noindent {\em if the MAOS calculation returns at least a complex solution, but only for 
one of the partitions, then this partition should be regarded as the wrong one.}
\medskip

\noindent
In the following we will refer to this statement as the complex-MAOS-solution criterion.
This requirement has a probability of mistakenly discarding a
correctly paired event as small as 5 per mil, as can easily be deduced
from the above items. Note that this criterion is 
already taken into account in the 86.8\% efficiency figure reported in 
eq. (\ref{eq:deltamWInequality}) and should be regarded as part and parcel 
of the $\Delta T_4$ method.

The whole discussion between eq. (\ref{eq:deltamtInequality}) and the previous paragraph
is illustrated in the plots of fig.~\ref{fig:DeltaT3T4}, that report the distributions for 
$\Delta T_3(P_W,\,P_R)$ and $\Delta T_4(P_W,\,P_R)$ 
(we recall the reader that $\Delta T$ is defined in eq. (\ref{eq:Tdiff})). In particular,
the efficiency figures in eqs. (\ref{eq:deltamtInequality}) and (\ref{eq:deltamWInequality})
correspond to the solid-line histograms, namely those before the $M_{T}^{t \bar{t}}(0)$ cut 
to be introduced in sec. \ref{sec:cut} (see figure legend).

A careful reader may have noted that, in the $\Delta T_4$ panel, the percentage of events 
in the positive semiaxis looks, and in fact is, smaller that the corresponding efficiency 
(percentage of correctly partitioned events) indicated in the legend. This is especially 
true for the histogram after the $M_T^{t \bar t}(0)$ cut.
The reason is the fact that the $T_4$-method efficiency is calculated as the sum of
the $\Delta T_4 > 0$ criterion {\em and}, when applicable, of the complex-MAOS-solution
criterion, that by itself belongs to the $T_4$-method, as already mentioned.
Schematically, the total $T_4$-method efficiency, $\epsilon_{T_4}$, splits up as follows:
\beq
\label{eq:T4efficiency}
\epsilon_{T_4} ~=~ C \cdot \epsilon_C + R \cdot \epsilon_R~,
\eeq
where $C$ is the percentage of events with complex solutions, $R$ is the percentage of events
free of complex solutions ($C + R = $100\%), $\epsilon_C$ denotes the efficiency of the 
complex-MAOS-solutions method and $\epsilon_R$ the efficiency of the $\Delta T_4 > 0$
method. For the events passing the $M_T^{t \bar t}(0)$ cut, we found $R = 31\%$ and 
$\epsilon_R \simeq 90\%$. The $R$ subset is the only one displayed in the dashed histogram
on the right panel of fig. \ref{fig:DeltaT3T4}, and the efficiency $\epsilon_R$ is the
percentage of this histogram lying in the positive semiaxis. More importantly for the
overall method efficiency $\epsilon_{T_4}$ in eq. (\ref{eq:T4efficiency}) is to note
that $C = 1 - R = 69\%$ and that $\epsilon_C \simeq 99\%$, to be compared with $C = 38.5\%$
(and $\epsilon_C$ basically the same) before the cut. These figures demonstrate that, after 
the cut, wrongly partitioned events tend to display complex MAOS solutions way more often 
than before the cut. This behavior is actually expected: if the $T_4$ variable, constructed 
with wrongly partitioned kinematics, may produce complex solutions, they will tend to 
proliferate if the kinematics gets more boosted, as is the case after the cut. 
The bottom line is that, from the point of view of the overall $T_4$-method efficiency, an 
increase in the number of events with complex solutions $C$ is an advantage, because 
the complex-MAOS-solution criterion has by itself an efficiency $\epsilon_C$ of nearly 
100\% of picking up the correct partition.

To conclude this section, we note explicitly that the previous discussion refers to events 
at parton level, as assumed throughout this paper. Of course, effects such as momentum smearing 
may well destabilize the above conclusions. However, it seems reasonable to believe that the 
main effect of smearing will not be to have complex solutions migrate from the wrong to the 
right partition, but rather to increase the number of events where {\em both} the right and 
wrong partitions consist of only complex solutions. Such events represented only 0.6\% of 
our total events (see point (b) above), and we simply did not use the $T_4$ variable for 
such events. It is possible that this fraction of events increases, maybe even substantially,
when including mismeasurement effects. Rather than simply disposing of the $T_4$ variable for 
these events, one may instead consider introducing some more general complex-solution 
criterion than the one advocated in this section. One such criterion may be to choose 
as the true partition the one whose MAOS solution has, for instance, the largest real part. 
It is worth mentioning that the issue of complex solutions in event reconstruction has been 
recently discussed in detail in ref. \cite{Gripaios:2011jm}, and the argument in the present 
paragraph is inspired by the main messages in that paper.
We leave the question of the utility and efficiency of our thus-reformulated complex-MAOS-solution 
criterion to forthcoming work \cite{futureWork} where smearing effects will be taken into
account.

\section{Improving the method efficiency}\label{sec:improvingmethod}

\subsection{Correlations and combinations of the test variables}\label{sec:combination}

In the previous section we have shown that each of the introduced test variables $T_i$ is 
able to find the correct final-state partition in $\gtrsim 80$\% of the total events 
that pass the basic selection cuts.
Already at the end of section \ref{sec:example} we have emphasized that the overall method 
efficiency may be systematically improved by combining the $T_i$ and/or by introducing 
appropriate kinematic cuts. We will now address these two possibilities in turn.

\begin{figure}[t!]
  \begin{center}
    \includegraphics[width=0.40\textwidth]{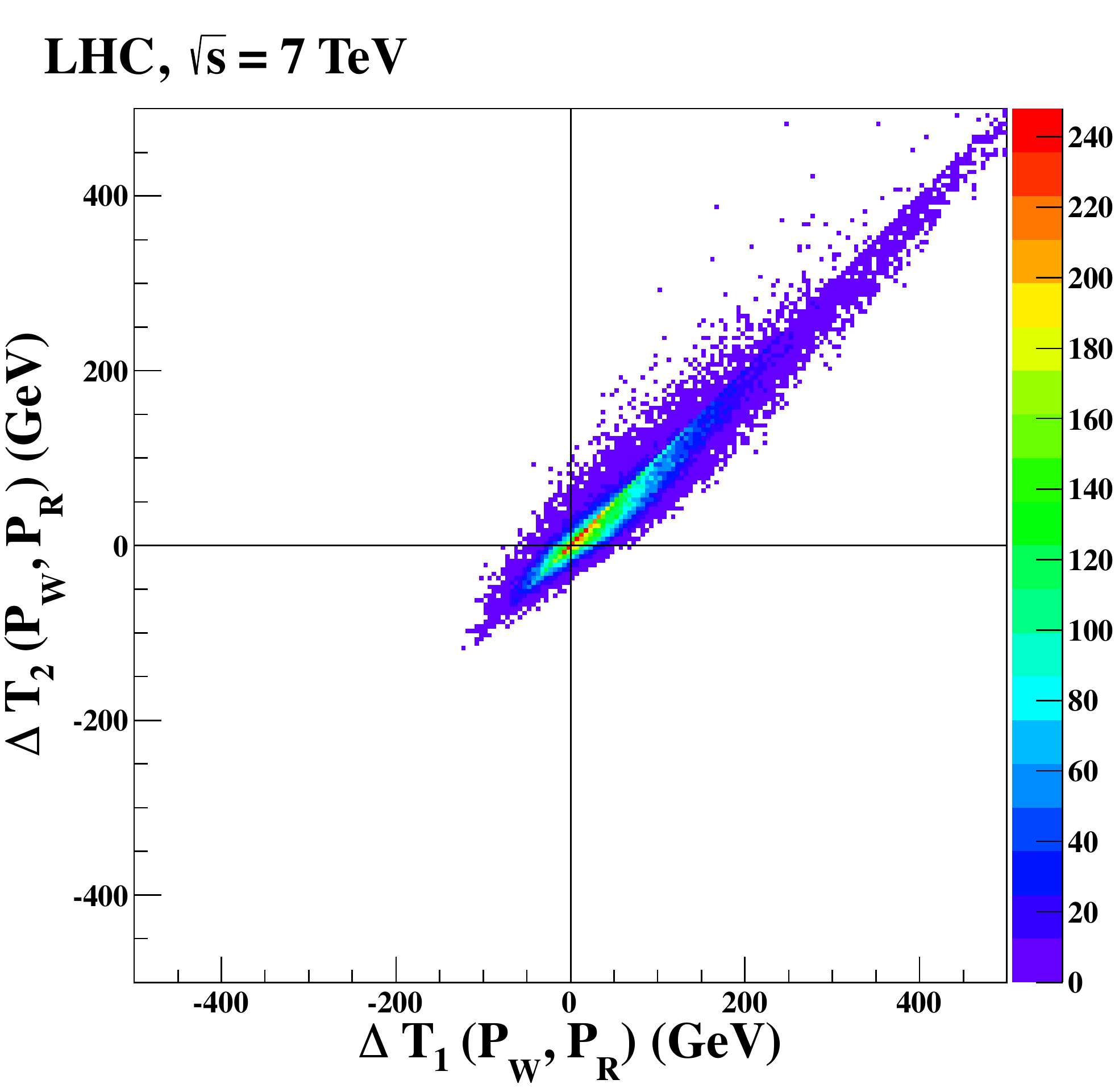}
    \includegraphics[width=0.40\textwidth]{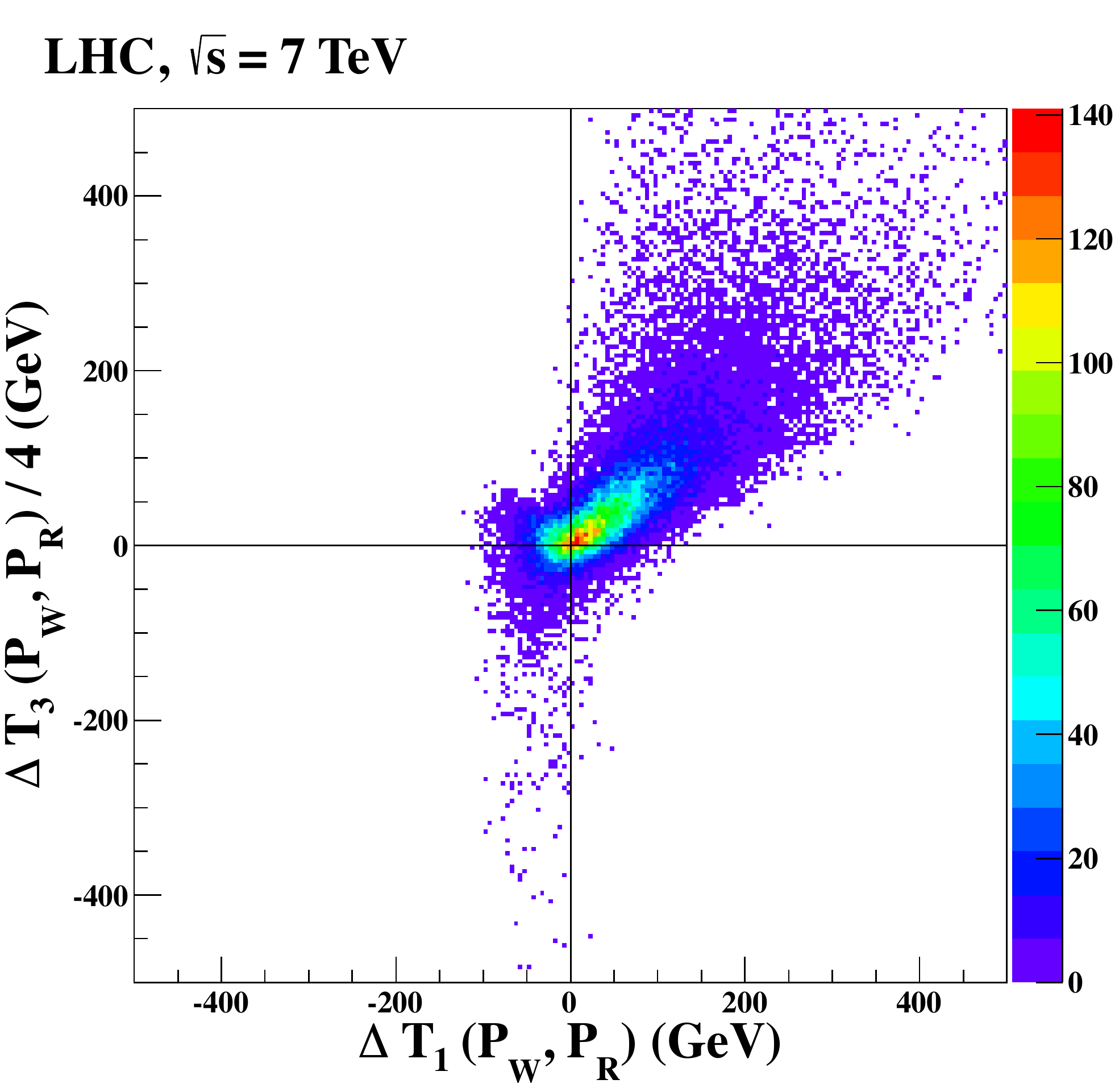}
    \includegraphics[width=0.40\textwidth]{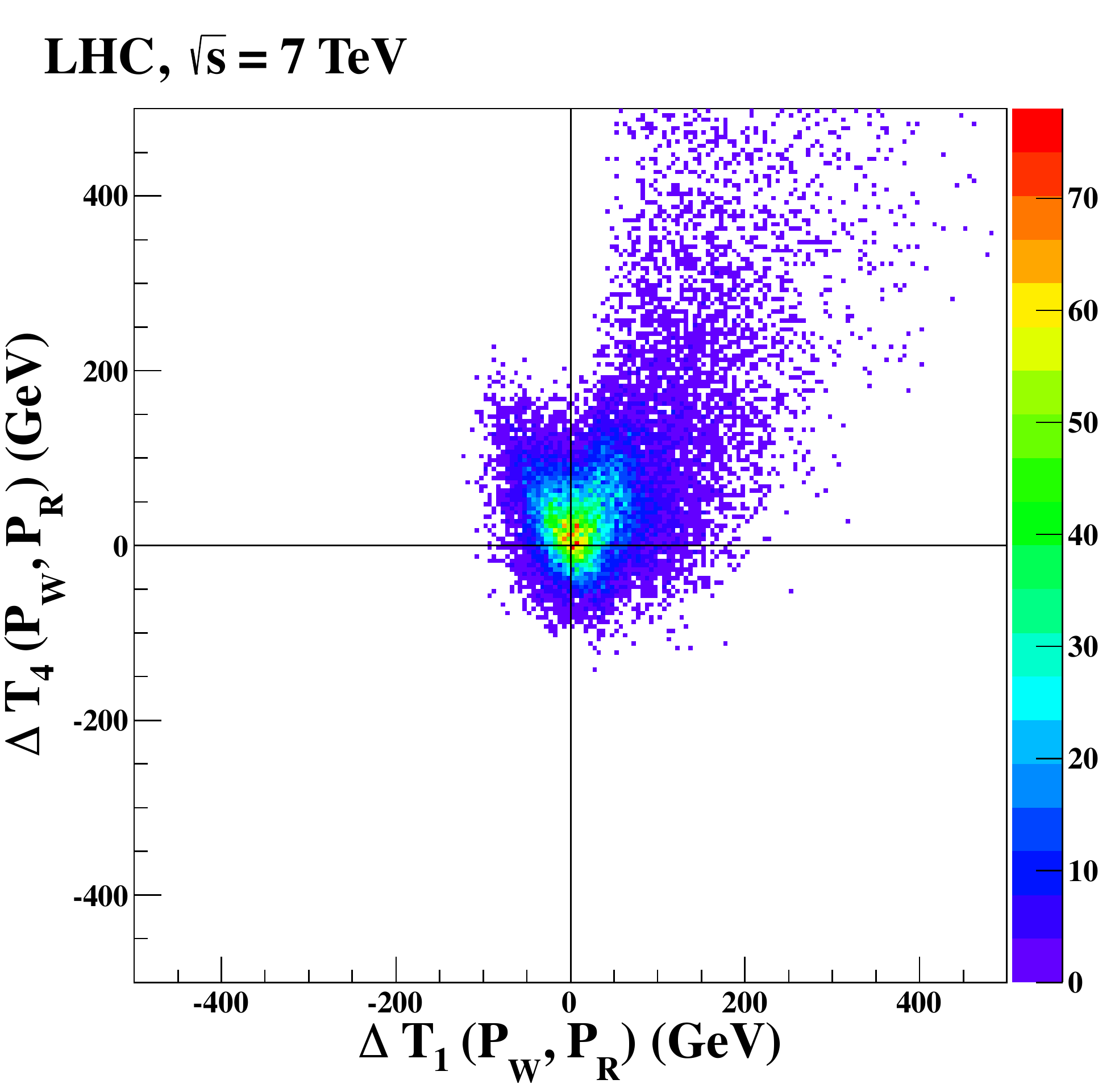}
    \includegraphics[width=0.40\textwidth]{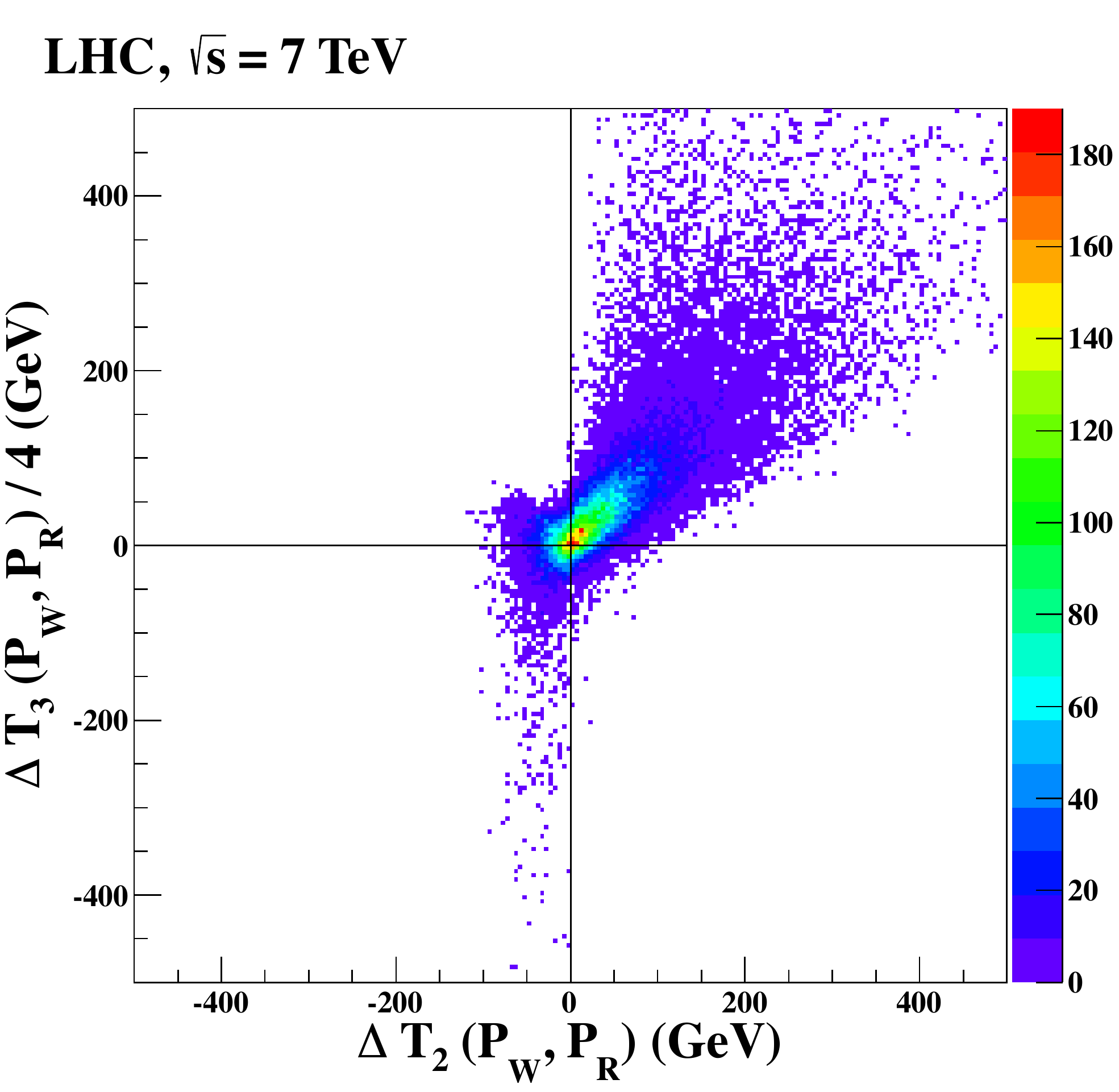}
    \includegraphics[width=0.40\textwidth]{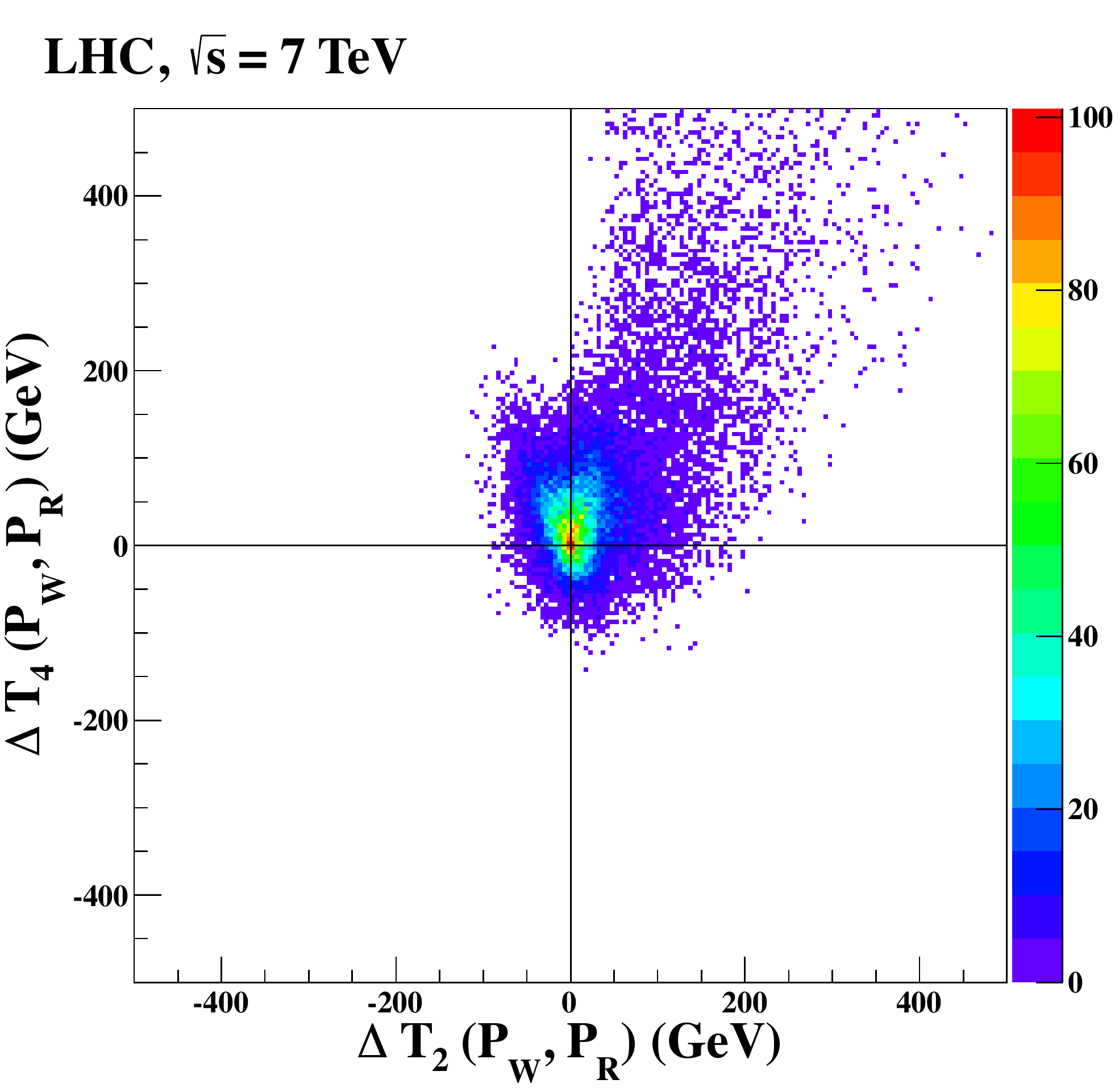}
    \includegraphics[width=0.40\textwidth]{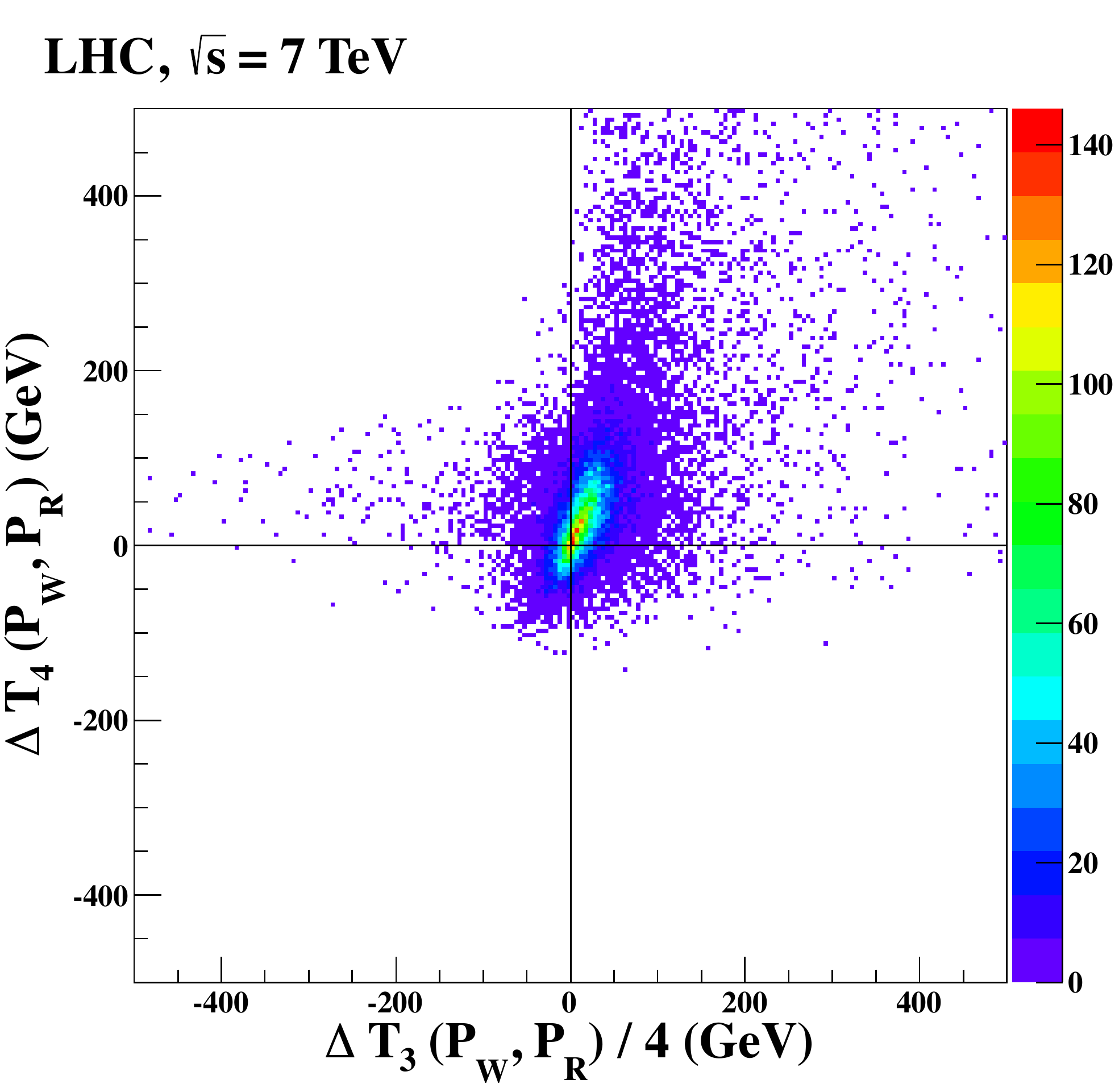}
  \end{center}
  \caption{Plots of the correlations between $\Delta T_i (P_W,\,P_R)$
    and $\Delta T_j(P_W,\,P_R)$. The color code refers to the number of
    events in each plot pixel.}
  \label{fig:corr}
\end{figure}

First of all, the possibility that combining the $T_i$ does indeed allow to increase the
overall method efficiency relies on the $T_i$ being as weakly correlated as possible. 
Hence, we should first address the question of how large these correlations are. For the
reader's convenience, we recall that our $T_i$ are defined in eqs. (\ref{eq:T1def}), 
(\ref{eq:T2def}), (\ref{eq:T3def}) and (\ref{eq:T4def}) and that we also introduced a
{\em complex-MAOS-solution} criterion at the end of sec. \ref{sec:T34}.
Two-dimensional plots of the $\Delta T_i(P_W,\,P_R) - \Delta T_j(P_W,\,P_R)$ correlations 
are shown in fig. \ref{fig:corr}. These plots do not include events with complex solutions.

The following comments are in order. First, it is apparent that $T_1$ and $T_2$ have a quite 
strong correlation. This is because the $m_{b\ell}$ value is directly used in the calculation 
of $M_{T2}$, that is a monotonically increasing function of $m_{b\ell}$. Further, the
MAOS-related variables, $T_3$ and $T_4$, are expected to have some degree of correlation with 
respectively $T_1$ (related to $m_{b \ell}$) or $T_2$ (related to $M_{T2}^{t\bar{t}}$). 
Arguments in support of this statement go as follows. Concerning the $T_3 - T_1$ correlation, 
$m_{t}^{\rm maos}$ was defined as
$$
\left(m_t^{\rm maos}\right)^2 = 
m_{b\ell}^2 + 2 \left( \sqrt{(m_{b\ell})^2 + |{\bf p}_{b\ell}|^2} |
{\bf k}^{{\rm maos}\mbox{-}WW}|- {\bf p}_{b\ell} \cdot {\bf k}^{{\rm maos}\mbox{-}WW} \right),
$$
whence the dependence on $m_{b\ell}$ is apparent. With regards to the $T_4 - T_2$ correlation, 
recall instead that the $m_{b\ell}$ value was used when calculating $k^{\rm maos\mbox{-}t\bar{t}}$, 
in turn necessary for $m_W^{\rm maos}$.
However, in practice, these correlations turn out to be rather weak, and this occurs 
because of the nontrivial structure of the MAOS solutions. For example, if $M_{T2}^{t\bar{t}}$ 
is calculated from a balanced configuration, $m_t^{\rm maos}$ can also be written as
\bea
  \left(m_t^{\rm maos}\right)^2 &=& 
  m_{b\ell}^2 + 2 \left( \sqrt{(m_{b\ell})^2 + |{\bf p}_{T,b\ell}|^2} |{\bf k}_T|
  \cosh(\eta_{b\ell} - \eta_{\nu})
  - {\bf p}_{T,b\ell} \cdot {\bf k}_T \right) \nonumber\\
  &=& \left(M_{T2}^{t\bar{t}}\right)^2 + 
  2\sqrt{(m_{b\ell})^2 + |{\bf p}_{T,b\ell}|^2} |{\bf k}_T| \left[
    \cosh(\eta_{b\ell} - \eta_\nu) - 1\right] ,
\eea
for given ${\bf k}_T$ values. Therefore, we can regard each of $\{ T_3, T_4 \}$ vs. 
each of $\{ T_1, T_2\}$ as approximately uncorrelated.

In principle, one can merge the information from the various $T_i$ even in presence of 
correlations among them, e.g. by constructing a likelihood function (not to mention more 
sophisticated pattern-finding techniques such as neural networks). The exploration of optimal
ways of using the combined information from all the $T_i$ is of course beyond the scope of the
present paper. Besides, it should be emphasized that each of the introduced variables can 
actually be defined in a number of variations. One may adopt the approach of considering all 
these variations and using their information in a joint way, that properly takes into account
correlations. This approach would also offer a way to improve the method efficiency. We 
refrain from entertaining such approach in this paper, which is devoted to the discussion of
the main ideas behind each variable.

Here we limit ourselves to show that even a very simple combination of the test 
variables results in an appreciable improvement of the overall method
efficiency.
We choose $\{T_2,\,T_3,\,T_4\}$ as a set of independent test
variables.
Then, the arguably simplest algorithm to find the right partition by
using these variables in a combined test is as follows:
\begin{itemize}
  \item[0.] Calculate $T_2$, $T_3$ and $T_4$ for the two possible partitions: $P_1$ and $P_2$;
  \item[1.] If $T_1(P_i)> m_{b\ell}^{\rm max}$ or $T_2(P_i)>m_t$, regard $P_i$ as the wrong 
  partition;
  \item[2.] If $T_4(P_i)$ results in complex MAOS solutions, whereas no complex solutions are
  present for the other partition, regard $P_i$ as the wrong partition;
  \item[3.] If none of the above criteria is met, then choose $P_1$($P_2$) as the right 
  partition if the majority of $\Delta T_i(P_2,\,P_1) >0$ ($<0$).\footnote{%
      A more refined approach along the lines of point 3 
      is to construct a function of the $T_i$, e.g., $\prod_i T_i$ 
      or a linear combination of $T_i$'s, and choose one partition using 
      the criterion $f(T_i)(P_W) > f(T_i)(P_R)$. We found this kind of 
      approach to also improve efficiency. The likelihood function is just 
      one of such functions.}
\end{itemize}
Point 1 corresponds to the already mentioned, and well-known, observation that $m_{b\ell}$ 
and $M_{T2}^{t\bar{t}}$ have a physical upper bound at $m_{bl}^{\rm max}$ and $m_t$ respectively,
if the partition is the correct one.
Point 2 is the complex-MAOS-solution criterion enunciated at the end of sec. \ref{sec:T34}.
As also discussed there, this method selects the right partition for $98.4\%$ of the events with 
at least one complex solution.
According to our simulation, the combined algorithm 0-3 described above returns the right 
partition for 89\% of the full data set, which we find an already remarkable improvement over 
the single-variable efficiency.

\subsection{Improving by cuts on kinematic variables} \label{sec:cut}

Up to this point, we have not introduced any event selection cuts, except for the basic cuts 
described in sec.~\ref{sec:eventgeneration}, that, we checked, have only a marginal impact on the 
$T_i$ efficiencies. Although the previous section shows that a method efficiency of 90\% or even 
larger is arguably obtainable by just a wise combination of the $T_i$, we would like to show in 
this section that improvements are possible also by imposing appropriate kinematic variable cuts. 
This of course comes at the price of losing (some) event statistics.

First, we note that the property of the test variables of being larger when evaluated with the 
wrong partition becomes more marked as the final states are more energetic. Intuitively, the larger
the magnitude of the input kinematics, the more spread becomes the distribution of the variable, if
it is calculated with the wrong partition. On the other hand, the distribution calculated with the 
right partition enjoys the same physical endpoints independently, of course, of the input kinematics.
So, for a more boosted subset of events, one expects $\Delta T_i(P_W, P_R)$ to be larger than zero more 
often. To single out such event subset, one may enforce cuts on appropriate kinematic variables,
and parametrically improve the method efficiency as the cut gets stronger.

\begin{figure}
  \begin{center}
  \includegraphics[width=0.49\textwidth]{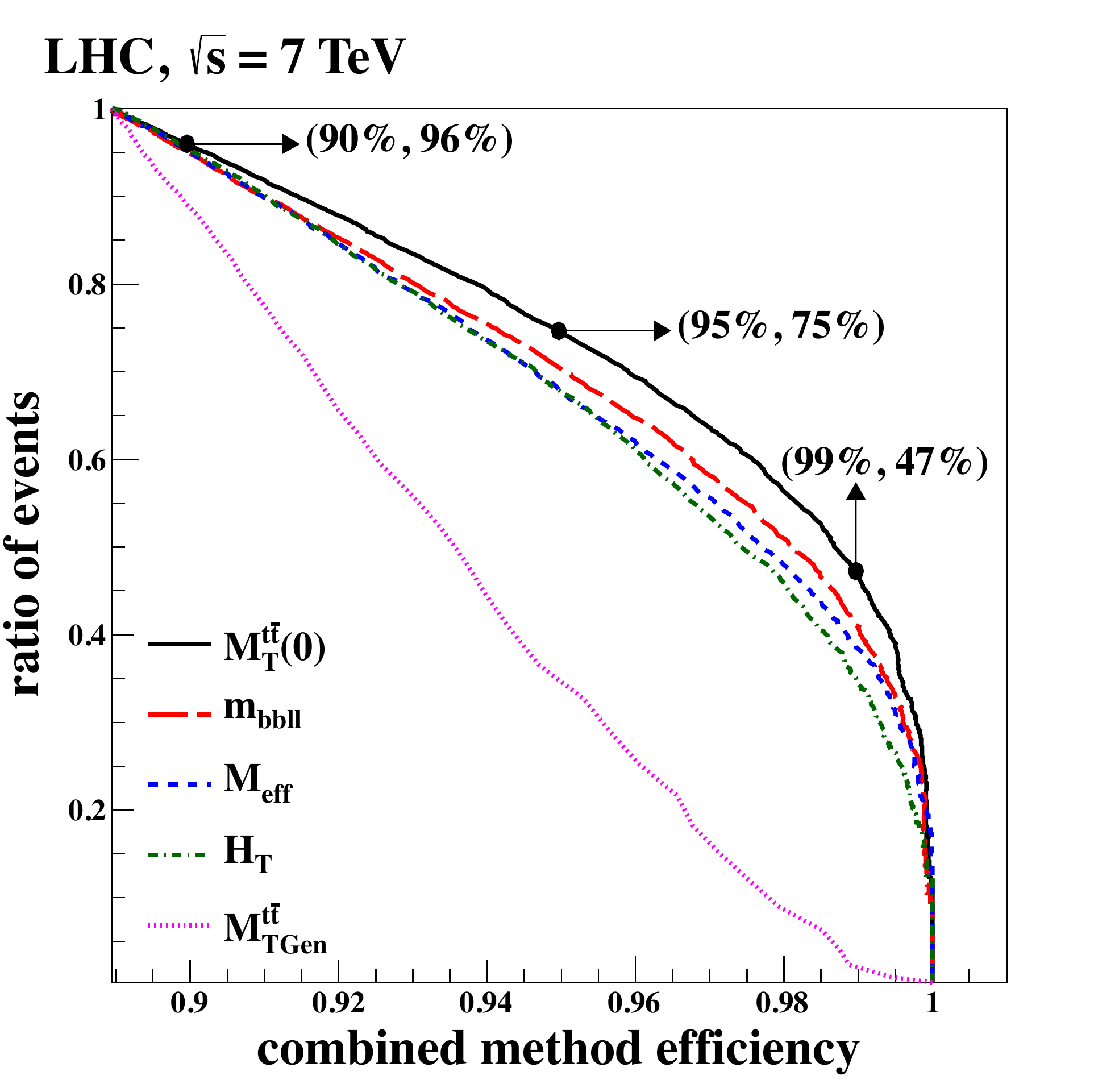}
  \includegraphics[width=0.49\textwidth]{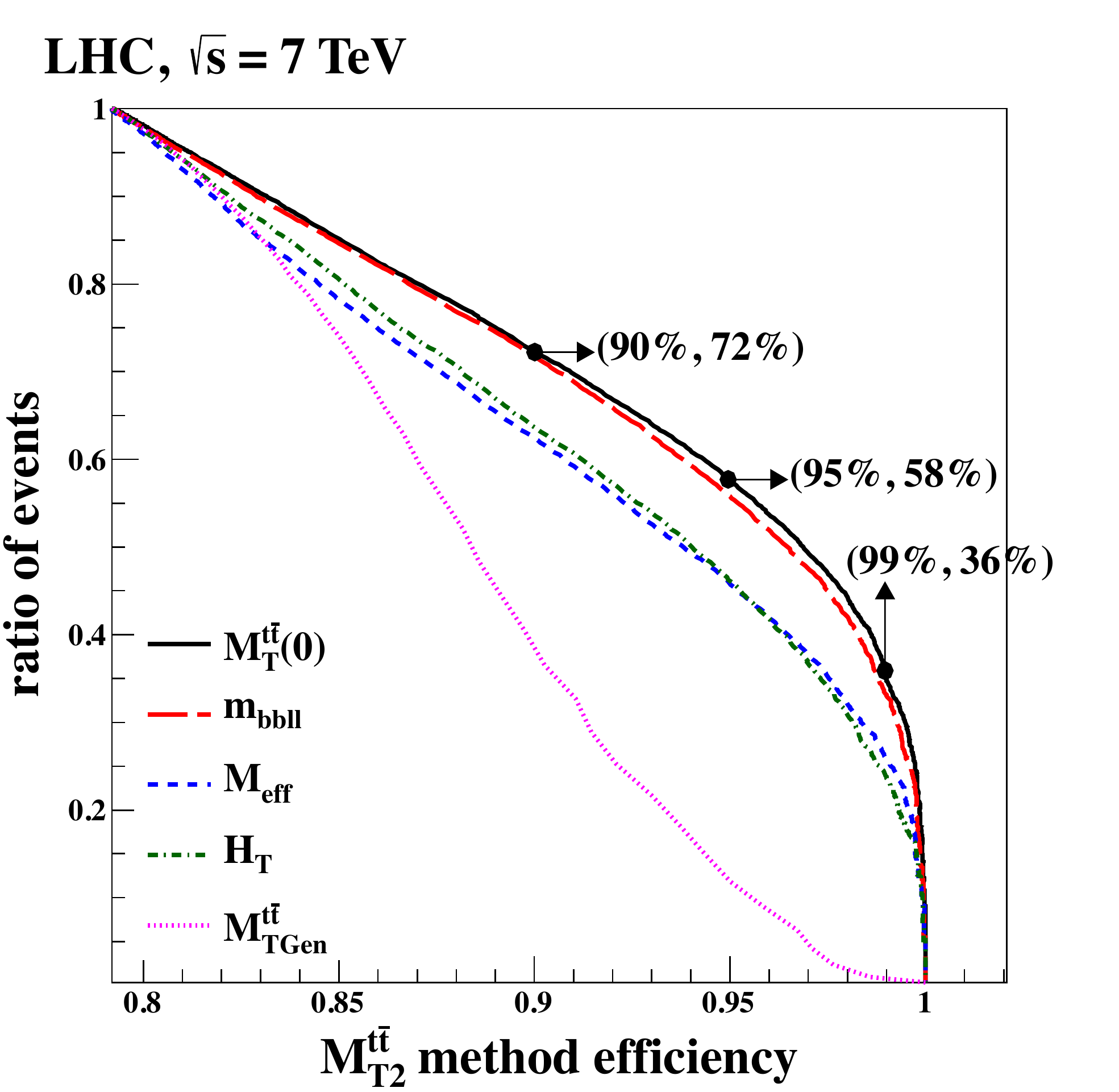}
  \caption{(Left panel)
  Combined method efficiency vs. loss in statistics with the inclusion of an inclusive cut. 
  The $x$-axis represents the efficiency of the combined method described in sec. 
  \ref{sec:combination} whereas the $y$-axis reports the number of events that passed
  one of the inclusive cuts, listed with different line/color codes in the legend. 
  Events are normalized to the event subset that passed the basic selection cuts discussed in 
  sec. \ref{sec:eventgeneration}.
  (Right panel)
  Same as the left panel, but for the $T_2$ variable only, rather than the combined method, 
  on the $x$-axis.}
  \label{fig:combine_cut}
  \end{center}
\end{figure}

For the dileptonic $t\bar{t}$ process, a well-known set of `partition-insensitive' kinematic 
variables --~which namely treat all the visible particles on an equal footing, hence they are 
free of combinatorial ambiguities~-- is represented by:
\begin{itemize}
  \item $M_T^{t\bar{t}}(0)$,
  \item $m_{V}^2 \equiv (p_{b_1} + p_{b_2} + p_{\ell_1} + p_{\ell_2})^2$,
  \item $M_{\rm eff} = \sum_{i} p_{{\bf i}T} + |{\bf \sla p}_T|$, 
  \item $H_T = \sum_{i} p_{{\bf i}T}$, 
    where $i = b_1,\,b_2,\,\ell_1,\,\ell_2$,
  \item $M_{T{\rm Gen}}^{t\bar{t}}$,
\end{itemize}
where $M_T^{t\bar{t}}(0)$ is the transverse mass of the full $t\bar{t}$ system with 
$m_{\nu\nu} = 0$~\cite{Park:2011uz,Barr:2009mx,Tovey:2010de},
\bea
  \left( M_T^{t\bar{t}}(0) \right)^2
  \equiv m_V^2 + 2 \left( \sqrt{|{\bf p}_T|^2 + m_V^2} |{\bf \sla p}_T|
  - {\bf p}_T \cdot {\bf \sla p}_T \right)
  \quad \mbox{with }{\bf p} \equiv \sum_{i=b,\,\ell} {\bf p}_i~,
\eea
and $M_{T{\rm Gen}}^{t\bar{t}}$ is the smallest value of $M_{T2}^{t\bar{t}}$ obtained over 
all possible partitions~\cite{Lester:2007fq}. One can impose these cuts in addition to the 
basic event selection cuts of sec.~\ref{sec:eventgeneration}. In the left panel of 
fig.~\ref{fig:combine_cut}, we show the flow of method efficiency vs. loss in statistics as 
one makes the cuts harder. See figure caption for full details.

The figure shows that the most effective cut variable is $M_T^{t\bar{t}}(0)$. One can reach
90\% method efficiency (i.e. get the right partition 90\% of the times) with just a 4\% loss
in event statistics. The efficiency rises to 99\% when roughly halving the statistics.
For the sake of completeness, we also mention that, for the quoted efficiencies of
90\%, 95\% and 99\%, the cut value is respectively $M_T^{t\bar{t}}(0) >$ 277, 343 and 404 GeV.

Because, for some event topologies to be described in the next section, the MAOS-based variables 
$T_3$ and $T_4$ are not usable, but the $M_{T2}$-based variable $T_2$ still is, we also show in 
the right panel of fig.~\ref{fig:combine_cut} a plot of efficiency gain vs. statistics loss 
for the $T_2$ variable alone.

\section{Generalizations of the method}\label{sec:generalization}

\subsection{Case of chain-assignment as well as ordering ambiguities}

From the discussion so far, our method of finding the right partition can be 
promptly generalized, in particular to topologies relevant for new physics scenarios. 
Among the most popular such event topologies is the one in eq. (\ref{eq:eventTopology}), 
that is, pair-production of heavy particles, decaying into a set of visible 
particles plus a pair of invisible particles (possibly, dark-matter candidates).
We here sketch this generalization, leaving full details to future work \cite{futureWork}.
For the sake of discussion, we here focus on identical decay chains, but the argument 
presented below can be extended also to the case of non-identical decay chains.

The first observation to make is that the decay of heavy parent particles can 
be in one step or multiple steps, depending on the mass spectrum and couplings of the 
new particles. The $T_1$ and $T_2$ variables of secs. \ref{sec:T1} and \ref{sec:T2} 
can be defined, and applied to find the right partition of the visible particles, 
even in the case of a one-step decay, e.g., gluino-pair production, followed by the
three-body decay $\tilde g \to qq\tilde\chi_1^0$ in $R$-parity conserving supersymmetric 
models.
\begin{figure}[t]
  \begin{center}
    \includegraphics[width=0.80\textwidth]{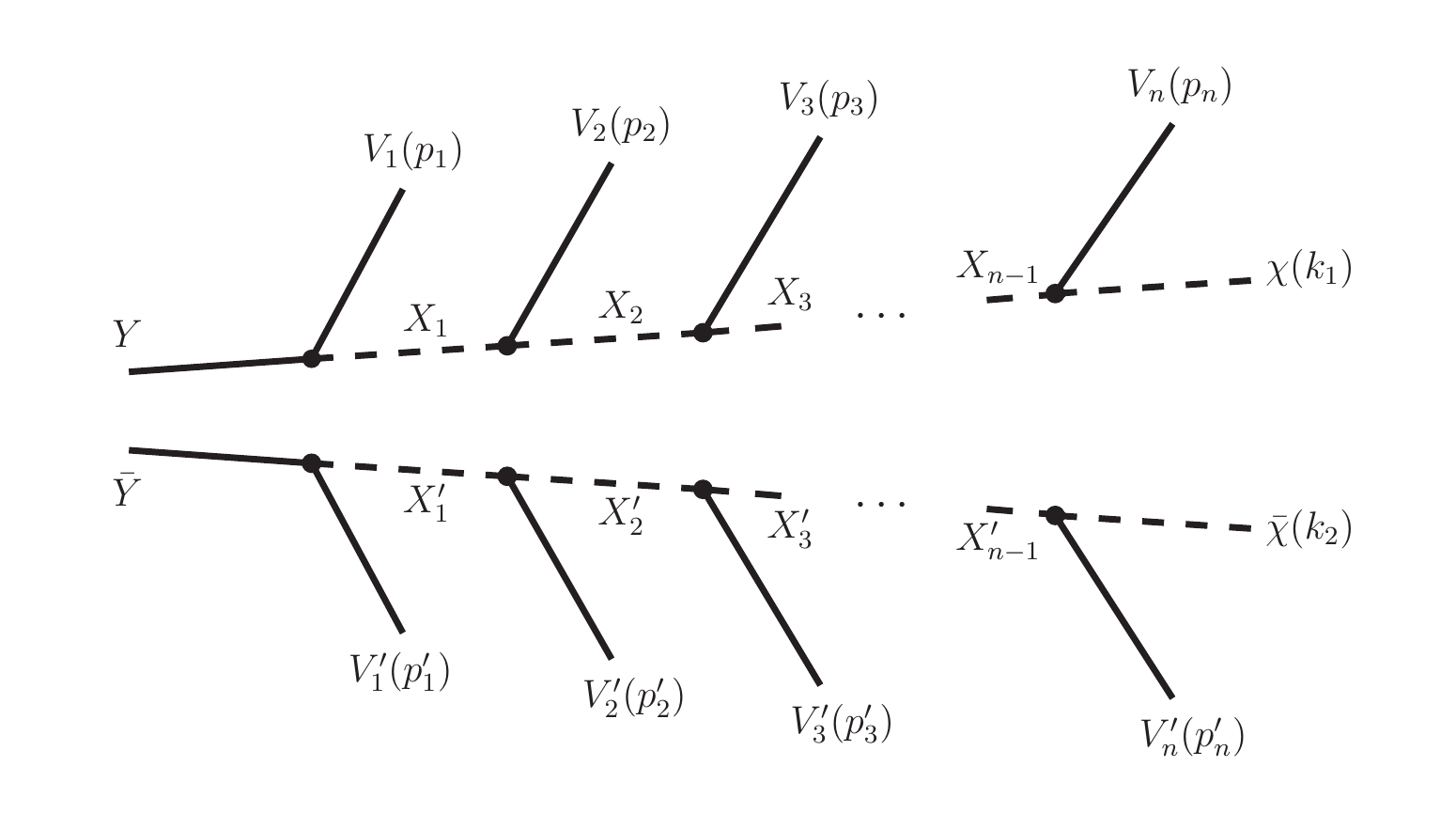}
    \caption{The decay process of eq. (\ref{eq:generalTopology}). The momenta $p_i$
    and $p'_i$ are assumed to be measurable, while $k_i$ give rise to missing transverse
    momentum.}
  \end{center}
  \label{fig:YYsketch}
\end{figure}
On the other hand, to utilize the MAOS-based variables $T_3$ and $T_4$, 
the decay process should be at least in two steps, as e.g. the dileptonic $t\bar{t}$ 
process is. For instance, if $m_{\tilde q} < m_{\tilde g}$, gluino-pair production 
will be dominantly followed by two-step decays, i.e., 
$\tilde g\to q\tilde q\to qq\tilde\chi_1^0$. In terms of event topology, the cascade 
decay processes of interest to this discussion can be generalized as
\begin{eqnarray}
  Y \bar{Y} &\to & 
  V_1(p_1) X_1 + V_1^\prime (p_1^\prime) X_1^\prime
  \to V_1(p_1) V_2(p_2) X_2 + 
  V_1^\prime (p_1^\prime) V_2^\prime (p_2^\prime) X_2^\prime
  \nonumber\\
  &\to & \cdots
  \to V_1 (p_1) \cdots V_n(p_n) \chi (k_1) +
  V_1^\prime (p_1^\prime) \cdots
  V_n^\prime (p_n^\prime) \bar\chi~(k_2)~,
  \label{eq:generalTopology}
\end{eqnarray}
for $m_Y > m_{X_1} > m_{X_2} > \cdots > m_\chi$ and
$m_{\bar Y} > m_{X_1^\prime} > m_{X_2^\prime} > \cdots > m_{\bar\chi}$. 
This decay is also sketched in fig. \ref{fig:YYsketch}. The rest of the
present discussion will also assume $m_{X_i} = m_{X_i^\prime}$, albeit this
requirement is renounceable with appropriate modifications of the method.

The second observation is that, in the context of this general event topology, 
there is, when partitioning the visible particles, an {\em additional} type of 
combinatorial ambiguity with respect to dileptonic $t\bar{t}$.
It is the {\em ordering} ambiguity, corresponding to the fact that 
$V_a^{(\prime)}$ is produced before $V_{a+1}^{(\prime)}$ ($a=1,\,\cdots,\,n-1$).
Of course, there is still, as in dileptonic $t \bar t$, the combinatorial ambiguity 
of {\em pairing} the visible particles into two sets, i.e., distinguishing 
$\{V_a\}$ from $\{V_a^\prime\}$. Then, the total number of partitions of the 
visible particles is
\begin{eqnarray}
  N = 
  (\mbox{number of pairings}) \times 
  (\mbox{number of orderings in one chain})^2 =
  \frac{_{2n}C_n}{2} \times (n!)^2~,
\end{eqnarray}
if the visible particles are not distinguishable among each other, as is the
case for jets (barring taggings etc.).
Then, one can calculate the MAOS invariant masses of $2n$ parent particles 
($Y,\,X_1,\,\cdots ,\,X_{n-1},\,\bar{Y},\,X_1^\prime,\cdots ,\,X_{n-1}^\prime$),
such that $n$ MAOS-based test variables are defined.
We will label them as $T_a^{\rm maos}$ ($a=0,\,\cdots ,\,n-1$),
for a given partition $P_j$ ($j=1,\,\cdots ,\,N$) and MAOS momenta $k^{\rm maos}$
(we recall that the MAOS determination of the invisible momenta consists 
in general of several solutions, collectively indicated as $k^{\rm maos}$). 
Thence, one can define:
\begin{eqnarray}
  T_a^{\rm maos}(P_j,\,k^{\rm maos}) \equiv \sum_{c=1,2;\,\alpha=+,-}
  |\Delta_{X_a} m_{X_a}^{\rm maos}(p_{a\,c},\,k_c^\alpha)|(P_j)~,
  \label{eq:Timaos}
\end{eqnarray}
where
\begin{eqnarray}
  \Delta_{X_a} m_{X_a}^{\rm maos}(p_{a\,c},k_c^\alpha)
  \equiv
  m_{X_a}^{\rm maos}(p_{a\,c},\,k_c^\alpha) - m_{X_a}
\end{eqnarray}
with $p_{a\,1} \equiv \sum_{j=a+1}^n p_j$, 
$p_{a\,2} \equiv \sum_{j=a+1}^n p_j^\prime$, and $X_0 = Y$.
These correspond to the $T_3$ and $T_4$ variables discussed in
sec.~\ref{sec:T34}. 
Actually, each $T_a^{\rm maos}(P_j,\,k^{\rm maos})$ can be obtained
in different ways, since the MAOS momenta $k^{\rm maos}$ can be
calculated from $_nC_2$ subsystems.
In order to account for them in a symmetric way, one may define
$T_a^{\rm maos}$ as
\begin{eqnarray}
  T_a^{\rm maos}(P_j) \equiv
  \sum_{\forall k^{\rm maos}} T_a^{\rm maos}(P_j,\,k^{\rm maos})~.
\end{eqnarray}
Each $T_a^{\rm maos}$ variable can then be used to decide on the right
partition: this partition is likely to have the smallest $T_a^{\rm maos}$
value than all the other partitions, irrespective of whether combinatorial
ambiguities concern the pairing or the ordering of the visible particle sets.

The $T_1$ and $T_2$ variables are also usable because one can calculate 
several subsystem $T_1$'s and $T_2$'s.
For example, in the event topology (\ref{eq:generalTopology}),
the subsystem $T_2$, or briefly $T_2^{\rm sub}$, can be defined as
$T_2^{Y\bar{Y}},\,T_2^{X_1X_1^\prime},\,\cdots ,\,
T_2^{X_{n-1} X_{n-1}^\prime}$, such that
$T_2^{Y\bar{Y}} \leq m_Y,\,T_2^{X_1 X_1^\prime} \leq m_{X_1}$, $\cdots$,
$T_2^{X_{n-1} X_{n-1}^\prime} \leq m_{X_{n-1}}$ for the right
partition. By sequentially testing the partitions in the order $Y\bar{Y}
\to \cdots \to X_{n-1} X_{n-1}^\prime$ systems, at each step selecting the 
partition with the smallest $T_2^{\rm sub}$, one can find the most likely 
partition for the full system.

We finally observe that the information from the above test variables can
be merged into a combined method in various ways. The simplest algorithm may be
\begin{itemize}
  \item[{\em (i)}] Calculate all the test variables as explained above.
    If $T_1^{X_a X_a^\prime} > m^{\rm max}_{V_{a+1}\cdots V_n} 
    = m_{V_{a+1}^\prime\cdots V_n^\prime}^{\rm max}$ 
    or $T_2^{X_a X_a^\prime} > m_{X_a}$ for $a=0,\,\cdots ,\,n-2$ 
    or there are more complex MAOS solutions in a given partition $P_j$ 
    than in any other partition, one should regard $P_j$ as the wrong one.
  \item[{\em (ii)}] If the above procedure fails to pick up one partition,
    one may select the right partition $P_j$ from the requirement that it 
    gives the smallest value of $T_i$ (with respect to the other partitions) 
    for the largest number of test variables.
\end{itemize}
We again emphasize that the efficiency of each $T_i$ and of the combined method 
can be further improved by imposing cuts on `partition-insensitive' variables,
as discussed in sec.~\ref{sec:cut}.

Concrete examples of this generalized method will be the gluino pair-production 
and decay discussed above, and the pair-production of the second lightest neutralino 
$\tilde\chi_2^0$, which decays to $\ell^+\ell^-\tilde\chi_1^0$. We will consider in 
more details these new physics processes for certain benchmark points in future 
work~\cite{futureWork}.

\subsection{Comments about the method's dependence on mass information}
\label{sec:depmasses}

We would like to conclude this section about generalizations with a few remarks
about the question of the method's dependence on the knowledge of the mass of the 
pair-produced states, $m_{Y}$, or that of the invisible daughters,
$m_\chi$.%
\footnote{We thank the Referee for triggering this discussion.}
As a first remark, we emphasize again that, of our proposed $T_i$, the variables
$T_1$ as well as $T_2$ do not use the information on $m_{Y}$ at all. On the other hand, 
this information appears necessary for constructing the MAOS-based variables $T_3$ and 
$T_4$, at least according to the way they were defined in sec. \ref{sec:T34}. In order 
to be able to use the variables $T_{3,4}$ at all, one would therefore seem to need 
$m_Y$ and $m_\chi$ to have been measured already.

It is worth observing that, in many scenarios of new physics, the mass measurements in 
question are not expected to be affected by 
combinatorial ambiguities. This holds in particular in the case of the topologies of 
interest for our method, where one can construct the event variable $M_{T2}$. In fact, 
if, in absence of combinatorial ambiguities, $m_Y$ may be estimated from the endpoint 
of $M_{T2}$, then, in presence of combinatorial ambiguities, the same estimate can be 
performed using $M_{T{\rm Gen}}$ \cite{Lester:2007fq} instead, i.e. the smallest 
$M_{T2}$ value for all possible partitions, which enjoys the same endpoint position 
as $M_{T2}$ constructed with the right partition.%
\footnote{The use of $M_{T{\rm Gen}}$ was also invoked in sec. \ref{sec:cut} as a 
partition-insensitive cut variable.}
In cases where the production cross-section is so small, or background contamination
so high, that the mass information cannot practically be extracted along the above lines,
one can consider using a modified version of our MAOS-based variables, wherein the 
parametric dependence on the true value of $m_Y$ is disposed of. To make an example, in 
place of the $T_3$ definition, given in eq. (\ref{eq:T3def}), one may instead use the 
quantity
\beq
\label{eq:T3alternative}
\tilde T_3(P_k) ~\equiv~ |m_t^{\maos}(\mbox{chain 1}) - m_t^{\maos}(\mbox{chain 2})|~,
\eeq
which should be close to zero if the partition is the correct one, whereas it can be 
much larger if the partition is wrong. Hence, again, $\Delta \tilde T_3(P_W, P_R)$ is 
expected to prefer positive values. We note explicitly that the notation in eq. 
(\ref{eq:T3alternative}) is, similarly as in eq. (\ref{eq:mtmaosInequality}), 
incomplete, in that the MAOS solution obtained from the on-shell relations 
(\ref{eq:MAOSonshellrelations}) is not unique (see discussion below those relations).
On the other hand, on the r.h.s. of eq. (\ref{eq:MAOSonshellrelations}), one does 
{\em not necessarily} have to choose the parent particle mass. One has in general 
the following choices \cite{Park:2011uz}
  \begin{itemize}
  \item MAOS1: $m_Y$ \cite{Cho:2008tj}
  \item MAOS2: $M_{T2} = \Min_{\vec p_{\chi_{1}, T} + \vec p_{\chi_{2}, T} = \sla{\vec{p}}_{T}}
  ~\Max \{ M_T^{(1)}, M_T^{(2)} \}$
  \item MAOS3: $M_T^{(i)}$, with $i = 1,2$ labeling the decay chain.
  \end{itemize}
The advantage of MAOS3 is that it allows to have a {\em unique} MAOS solution for 
each event, thereby allowing use of the $\tilde T_3$ variable as defined in eq. 
(\ref{eq:T3alternative}). In fact, we found the performance of $\Delta \tilde T_3$ 
comparable to that of $\Delta T_3$, as defined in sec. \ref{sec:T34}. The reason why 
we defined our MAOS-based variables according to the MAOS1 scheme is mostly simplicity.

The other unknown mass besides $m_Y$ is $m_\chi$, upon which all the $T_i$ variables
bear dependence. While the necessity to input some $m_\chi$ value seems inescapable, 
we are actually quite confident that our method would perform rather well even in the case 
where the $m_\chi$ mass is yet to be measured at the moment of having to solve the 
combinatorial problem. We make the following considerations in support of this statement.
First, as already mentioned in the general $M_{T2}$ discussion in sec. \ref{sec:T2}, 
an estimate, perhaps an accurate one, of $m_\chi$ may be obtained from the $M_{T2}$
kink \cite{Cho:2007qv,Cho:2007dh,Gripaios:2007is,Barr:2007hy}, which however 
requires sufficient statistics. 
Even in absence of such statistics, and in presence of combinatorial ambiguities, 
the uppermost part of the $M_{T {\rm Gen}}(m_\chi)$ plot in the $m_\chi$ vs. $m_Y$ 
plane may provide a rough idea of the $m_\chi$ mass.
It is worth emphasizing that a rough input value for $m_\chi$ may well be sufficient 
in many scenarios. It is so at least in the case where $m_\chi \ll m_Y$: it can be 
shown in fact \cite{Kiwoon_unpub} that, in this limit, MAOS solutions have 
corrections going as $m_\chi^2 / m_Y^2$ and are, therefore, largely insensitive 
to the specific value chosen for $m_\chi$, so long as the mentioned mass hierarchy
holds at least approximately.

Clearly, all the issues mentioned in this section deserve further investigation, 
which is best carried out within specific new-physics benchmark points. This lies outside 
the scope of the present paper, mainly meant to propose our method and to apply it to the 
simplest yet relevant SM example we could think of, namely $t \bar t$ production. We will 
however return to all these issues in a forthcoming paper \cite{futureWork}.

\section{Conclusions and Outlook}\label{sec:conclusions}

In this paper, we have proposed a novel method to resolve combinatorial 
ambiguities in hadron collider events involving two invisible particles 
in the final state. This method is based on the kinematic variable $M_{T2}$ 
and on the MAOS reconstruction of invisible momenta,
that are reformulated as test variables $T_i$, namely testing the correct 
combination against the incorrect ones. The efficiency of each single $T_i$
in the determination of the correct combination can then be systematically improved
in two directions: by combining the information from the different $T_i$ and by
introducing further cuts on suitable, partition-insensitive, variables.
In this sense, our method is completely scalable.

All the above program is illustrated in the specific, and {\em per se} interesting
example of top anti-top production, followed by a leptonic decay of the $W$ on 
both sides. We however emphasize that, by construction, our method is also directly 
applicable to many topologies of interest for new physics, in particular events 
producing a pair of undetected particles, that are potential dark-matter candidates. 

Our method opens a whole spectrum of generalizations and by-product issues, on which 
work is in progress \cite{futureWork}. The most urgent include:
\begin{itemize}
  \item Addressing whether and how the method is affected by 
  the inclusion of effects such as QCD radiation, hadronization and 
  signal degradation due to detector effects. The main challenge
  of this problem is that, after hadronization, the assignment of
  a `parton' to a given decay chain or to a certain position within
  the decay chain is no more well-defined. As such, it requires
  separate thought.
  \item Generalizing the method to the combinatorial uncertainty due to
  the {\em ordering} of the visible particles in the decay chain, 
  uncertainty which is not present in dileptonic $t \bar t$.
\end{itemize}

\acknowledgments
KC and DG gratefully acknowledge the hospitality of Universidad
Aut\'onoma de Madrid and CSIC, as well as of CERN (the TH - LPCC Summer Institute 
on LHC Physics) where part of this work was carried out.
We would like to thank W.~S.~Cho and Y.~G.~Kim for the discussion that laid
the foundation for this work. KC is supported by the KRF Grants funded 
by the Korean Government (KRF-2008-314-C00064 and KRF-2007-341-C00010) 
and the KOSEF Grant funded by the Korean Government (No. 2009-0080844).
The work of DG is supported by the EU Marie Curie IEF Grant no. 
PIEF-GA-2009-251871.
The work of CBP is partially supported by the grants FPA2010-17747, 
Consolider-CPAN (CSD2007-00042) from the MICINN, HEPHACOS-S2009/ESP1473 
from the C.~A.~de Madrid and the contract ``UNILHC'' PITN-GA-2009-237920 
of the European Commission.

\bibliographystyle{JHEP}
\bibliography{ttbar_MAOS}

\end{document}